\documentclass[12pt]{elsarticle}
\usepackage{graphicx} 
\usepackage{amsfonts} 
\usepackage{amssymb}
\usepackage{enumerate}
\usepackage{url}
\usepackage{rotating}
\usepackage{graphicx}
\usepackage{multicol}
\usepackage{dsfont}
\usepackage{caption}
\usepackage{amsmath, bm}
\usepackage{subcaption}
\usepackage{natbib}
\usepackage[]{algorithm2e}
\usepackage{multirow}
\usepackage{hhline}
\usepackage[export]{adjustbox}
\usepackage{algorithm2e}
\RestyleAlgo{ruled}
\usepackage{float}
\usepackage{color}

\usepackage[a4paper, left=3cm, right=3cm, top=3cm, bottom=3cm]{geometry}

\begin{document}
\begin{frontmatter}

\title{Effective toughness estimation by FFT based phase field fracture: application to composites and polycrystals}

\author[upm]{Pedro Aranda}
\ead{pedro.aranda@usach.cl}

\author[upm,imdea]{Javier Segurado}\corref{cor1}
\ead{javier.segurado@upm.es}
\address[upm]{Departamento de Ciencia de Materiales, Universidad Politécnica de Madrid, C/ Profesor Aranguren 3, 28040 - Madrid, Spain}
\address[imdea]{IMDEA Materials Institute, 28906, Getafe, Madrid, Spain}

\date{\today}

\begin{abstract}
An estimate of the effective toughness of heterogeneous materials is proposed based on the Phase Field Fracture model implemented in an FFT homogenization solver. The estimate is based on the simulation of the deformation of representative volume elements of the microstructure, controlled by a constant energy dissipation rate using an arc-length type control. The definition of the toughness corresponds to the total energy dissipated after the total fracture of the RVE ---which can be accurately obtained thanks to the dissipation control--- divided by the RVE transverse area (length in 2D). The proposed estimate accounts for both the effect of heterogeneity in toughness and elastic response on the overall fracture energy and allows as well to account for phases with anisotropic elastic and fracture response (fracture by cleavage). To improve toughness predictions, crack-tip enrichment is used to model initial cracks. The method is applied to obtain the effective toughness of composites and elastic polycrystals in a series of examples. In the two types of materials, it is found that both heterogeneity in elastic response and fracture energy contribute to increase the effective toughness. Microscopically, it is found that toughening mechanisms are related to the passage of the crack through tougher phases and deviation of the crack path. It is also found that the latter is the controlling mechanism for cases with marked heterogeneity and high anisotropy, eventually provoking toughening saturation for sufficiently high values of heterogeneity or anisotropy.
\end{abstract}
\end{frontmatter}

\section{Introduction}
\label{sec:Int}
The resistance to crack propagation in any material, also known as fracture toughness, is controlled by its nature and microstructure, which include phase arrangement, presence of inclusions and others. The estimation of the microstructure influence on the overall toughness has a significant engineering relevance, enabling the design of microstructures that maximize fracture resistance.

Similarly to classic homogenization concepts such as effective stiffness or effective conductivity, \emph{effective toughness} can be defined as an overall value of toughness that characterizes the fracture of a heterogeneous material at the macroscale as a function of the constituent fracture energy and microstructure \cite{rose1986effective, braides1996homogenization}. Thus, micromechanical models in which microstructure is explicitly considered and mechanical microscopic fields are resolved will be the framework to obtain this effective response from the microstructure.

Early works consider this problem using analytical approaches, showing how the macroscopic stress intensity factor $K_c$ is modified due to the presence of microcracks \cite{rose1986effective} or by the differences in the phase properties at the microscale \cite{huajian1991fracture}. In \cite{roux2003effective} the effective measure is obtained by averaging the crack advance in a media with heterogeneous local $K_c$. A more precise estimation of the effect of the microstructure on the overall toughness can be obtained using numerical methods for solving the microscopic problem. Different effective toughness definitions have been used to motivate the microscopic simulations.

Braides et. al. \cite{braides1996homogenization} defined the effective toughness in the homogenization context as the minimum energy released by a crack in microstructural heterogeneous domains, assuming a full scale separation. In this definition, the microscopic elastic response is homogeneous, considering that under full scale separation, such optimization is independent of the local stress distribution. This definition was used by \cite{schneider2020fft, ernesti2021fast, ernesti2022computing} who translated the optimal crack path search into a variational problem to find a diffuse crack definition with minimum energy, where optimal microscopic crack paths are explored on representative volume elements (RVEs) with increasing sizes. The advantages and limitations of this variational definition have recently been analyzed in \cite{MICHEL2022104889}. The effective toughness is formulated as the spatial average of the local toughness along the path. An algorithm based on a primal-dual method (PM) and an alternating direction method of multipliers (ADMM) implemented in a FFT based scheme was proposed to solve the problem. Similar ideas have been explored in \cite{lebihain2020effective, lebihain2021effective} where an stochastic/perturbative method based in LEFM is used, considering weighted states of interaction between phases with high local toughness contrast.  

Alternatively, other authors \cite{li2013prediction} and \cite{hossain2014effective} rely on simulations including crack propagation on RVEs including a heterogeneous distribution of toughness and stiffness to estimate the effective toughness based on the value of a J-integral on the full RVE. Those works show that stiffness and toughness heterogeneities act as toughening mechanisms by deflecting or arresting microscopic cracks, which can be captured in J-integral estimations. This leads to definitions of effective toughness such as the one in \cite{li2013prediction} where a time-averaged value of the macroscopic J-integral (when fracture stabilizes) is taken as the effective crack resistance. Hossein et.al. \cite{hossain2014effective} consider the maximum value of the macro-J-integral as the effective toughness, under the argument that micro-cracks have to overcome that value at some point in order to complete the fracture process. These works require the full simulation of the fracture process at the microscale, and \cite{hossain2014effective} base these simulations on the Phase Field Fracture model in FEM (FEM-PFF). 

PFF model has gained strong acceptance for fracture simulations in the last years, being proposed primarily in \cite{FRANCFORTMARIGOJMPS1998, ambrosio1992approximation} and further developed by Miehe et.al in \cite{MieheIJNME2010} for elastic/brittle materials. The model is based on a variational approach in which cracks are represented as continuous fields with a high level of concentration in the position in which an actual crack would exist. This method presents advantages in estimating effective fracture toughness, as the elastic energy and fracture dissipation functionals of the material can be calculated directly during the fracture process \cite{na2017effects, hossain2014effective}. The model can consider heterogeneous microstructures as input, using local variations of $G_c$ and stiffness. Since the crack-growth driving force is an energy functional dependent on the local strain state, the microstructure can strongly influence the microcrack path and, as a result, the macroscopic estimate of $G_c$ \cite{hossain2014effective}.

One of the differences between the works of \cite{hossain2014effective} and \cite{schneider2020fft} is that the latter uses a highly optimized FFT-based solver, in contrast to standard FE simulations used in \cite{hossain2014effective}. FFT-based solvers are a well-established alternative to FEM simulations for micromechanical problems, thanks to their high computational efficiency and the periodic nature of its solutions that extend to all the fields considered in the problem \cite{lucarini2021fft}, including the PFF variable \cite{MA2020}, without the need of additional constrains. These methods allow the use of very complex microstructures, encapsulated in RVEs with very fine discretizations, since the method manages to scale the computational cost of $n$ degrees of freedom with a $n log(n)$ behavior. Damage and fracture approaches at the microscale have been proposed for FFT solvers, with the former based mainly on non-local continuum damage models for composites \cite{MAGRI2021113759,LI2025117854} or ductile polycrystals \cite{cocke2023implementation,pokharel2012quantifying}. The latter includes implementations of the above-mentioned PFF model, being used for the study of fracture in linear elasto-brittle materials \cite{chen2019fft, ernesti2019fft, Ernesti2020, aranda2025crack}, brittle strongly anisotropic materials \cite{MA2020}, elastoplastic materials with crystal plasticity \cite{cui2023applications, lucarini2023fft} or coupled with diffusion to study chemically assisted cracking \cite{zarzoso2025fft} . 

An effective measure of macroscopic toughness can be obtained by solving a full micromechanical simulation including PFF as in \cite{li2013prediction,hossain2014effective}, but using a definition of the macroscopic toughness based on the total energy released by the microscopic crack propagation. A challenge of this approach is how to measure the fracture energy dissipated in the RVE. Potentially, this value could be obtained in a PFF simulation under strain control as the value of the fracture functional at the final fracture stage \cite{na2017effects}. However, the unstable propagation is normally surpassed by the use of staggered algorithms, which makes uncertain the uniqueness of the final crack obtained. Moreover, since PFF models usually contain some non-variational elements (e.g. history dependence), in an unstable propagation the total fracture energy obtained using a staggered approach might not be unique and be different from the dissipation following a stable load path.

In our previous work \cite{aranda2025crack}, a PFF-FFT scheme with a crack-growth control was developed, allowing calculations over a \emph{snap-back} mechanical response in elastic brittle materials. This makes it possible to compute the total external energy in a controlled fracture process for a specific heterogeneous RVE even in cases where crack growth is unstable under a macroscopic strain/stress control. Under the assumption that all energy has to be dissipated in fracture, this measure can be used as an effective toughness estimate, similar to \cite{schneider2020fft}, but allowing one to consider the effect of microscopic strain distribution due to heterogeneity on the resulting crack path.

The effective toughness estimate based on PFF-FFT and the crack length control was proposed in \cite{aranda2025crack} for very simple cases of laminate microstructures with heterogeneity in stiffness and toughness, and transverse fracture of LFRC assuming a homogeneous mechanical response \cite{aranda2025crack}. Nevertheless, a systematic analysis of the homogenization of toughness using this estimate is missing, as well as its application to study fracture on materials at microscale considering highly heterogeneous and anisotropic media.

In this work, the effective toughness estimate based on PFF-FFT and dissipation energy control is extended to microstructural media with strong heterogeneous phases, elastic anisotropic mechanical behavior and marked damage anisotropy (cleavage). This framework will then be used to analyze the effective fracture toughness of composite materials and elastic polycrystals as a function of the stiffness and local toughness ratios of their constituents. In section 2, the mathematical formulation of the PFF fracture model is reviewed, and the definition of the effective fracture toughness estimate is proposed. Section 3 includes the numerical implementation of the model in a FFT-based solver, with the simulation conditions required to carry out simulations and the toughness measurement. Section 4 presents the geometrical characteristics and simulation conditions, as well as the results obtained for the different studies of effective toughness evolution.

 
\section{Models}\label{sec:MODELS}

\subsection{Phase Field Fracture model in homogenization simulations}\label{sec:classicPFF}
Phase Field fracture aims at modeling crack growth following the Griffith's energy criterion for brittle elastic materials. The model has been widely developed in the literature \cite{MieheIJNME2010,delorenzis15,onate2017advances} for different materials. This section provides a brief summary of the main aspects of the PFF model relevant to this study and the control technique proposed in \cite{aranda2025crack}.

\begin{figure}[htbp]
\centering
\includegraphics[width=100mm]{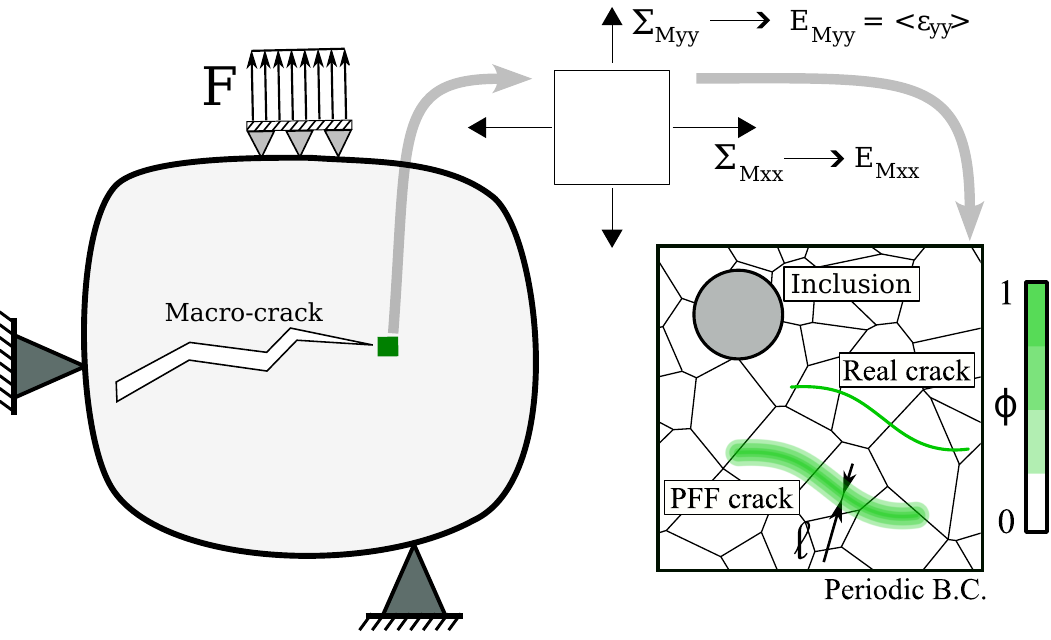}
\caption{\centering{Damage Phase Field representation of the crack.}}
\label{fig:PFFscheme}
\end{figure}

The model considers the representation of a discrete crack using a scalar field called damage $\phi$, which represents the presence of two phases: void ($\phi=1$) and pristine material ($\phi=0$). The interface between them is modeled to be continuous and smeared, and its size is controlled by a characteristic length $\ell$, allowing to define a crack as in Fig.\ref{fig:PFFscheme}.

Fracture propagation can be represented by a variational scheme, involving the strain field $\boldsymbol{\varepsilon}:\Omega \rightarrow \mathbb{R}^{3}\otimes\mathbb{R}^{3}$ and the damage field $\phi:\Omega \rightarrow \mathbb{R}$. Using the principle of virtual power, the power of external loads $\dot{\Pi}_{M}$ is balanced with the power of the internal energy $\Psi$ as
\begin{equation}
\label{eq:Free_Ener}
\dot{\Psi}(\vec{u},\phi) = \dot{\Pi}_{M} \hspace{3mm}\Rightarrow \hspace{3mm} \dot{E}_{mec}(\boldsymbol{\varepsilon},\phi) + \dot{D}(\phi) - \dot{\Pi}_{M} = 0.
\end{equation}
\noindent The internal energy is composed by a fracture energy term $D$ and the elastic free energy $E_{mec}$, which is obtained by integrating the elastic energy density $\psi$
\begin{equation}
\label{eq:elas_Ener}
E_{mec} =  \int_{\Omega} \psi(\boldsymbol{\varepsilon},\phi) \mathrm{d}\Omega = \int_{\Omega} g(\phi)\psi_{o}(\boldsymbol{\varepsilon}) \mathrm{d}\Omega = \int_{\Omega} g(\phi) \frac{1}{2} \mathbb{C}_o:\boldsymbol{\varepsilon}:\boldsymbol{\varepsilon}  \mathrm{d}\Omega,
\end{equation}
\noindent where $\psi_{o}$ corresponds to the undamaged density energy, $g(\phi)$ is a damage dependent function representing energy degradation and $\mathbb{C}_0$ corresponds to the pristine fourth order isotropic stiffness tensor that can vary in space. 

In Eq.\eqref{eq:Free_Ener}, the dissipation functional can be defined as a function of damage, $D(\phi)$, and represents the energy used to fracture the material, which is coupled to the elastic energy through a degradation function $g(\phi)$ defined quadratically as $g(\phi) = (1-\phi)^2 - k$, where $0<k<<1 $. The functional includes Griffith's critical energy release rate $G_c$ as local toughness and a crack dissipation density $\gamma_{f}(\phi)$ is defined in this work with the AT2 model \cite{Ambrosio1990,ambrosio1992approximation} as 
\begin{equation}
\label{eq:CrackLenght}
D(\phi,\ell,G_c) = \int_{\Omega} G_c(x) \gamma_{f}(x,\phi) d\Omega = \int_{\Omega} G_c(x) \left( \frac{1}{2\ell}\phi^{2}+\frac{\ell}{2} \nabla\phi\cdot\nabla\phi \right)d\Omega.
\end{equation}
Note that $G_c(x)$ depends on the material point to consider heterogeneous materials with different local toughness. The dissipation density functional can similarly define the total surface of the actual crack $\Gamma$ as:
\begin{equation}
\label{eq:CrackLenghtreal}
\Gamma(\phi,\ell) = \int_{\Omega} \gamma_{f}(x,\phi) d\Omega = \int_{\Omega} \frac{1}{2\ell}\phi^{2}+\frac{\ell}{2} \nabla\phi\cdot\nabla\phi d\Omega.
\end{equation}
The strong form of the problem governing equations is obtained by differentiation with respect to time and integration by parts of Eq.\eqref{eq:Free_Ener} \cite{MieheIJNME2010}. The resulting system of PDEs is given in Eq.\eqref{eq:SystemDis0}, 
\begin{subequations}
\label{eq:SystemDis0}
\begin{eqnarray}
\nabla \cdot \boldsymbol{\sigma}= \nabla \cdot g(\phi)\boldsymbol{\sigma}_{o} = 0 \label{eq:SystemDis0a}\\
g'(\phi)\psi_{o}(\boldsymbol{\varepsilon}) + \frac{G_{c}}{\ell}\phi-  \nabla\cdot\left [ \ell G_{c}\nabla\phi \right] = 0 \label{eq:SystemDis0b}\\
\dot{\Pi}_{M} =\tau(t),\label{eq:SystemDis0c}
\end{eqnarray}
\end{subequations}
In the system, the first equation corresponds to the mechanical equilibrium defined with the undamaged stress $\boldsymbol{\sigma}_{o}$. The second equation is derived from the variation of the damage and corresponds to the Helmholtz equation, and the function $g'(\phi)$ is the derivative respect to $\phi$ of the degradation function is
\begin{equation}
\label{eq:desgast}
 g'(\phi) = -2(1-\phi).
\end{equation}
Eq. \eqref{eq:SystemDis0b} represents the  balance between the elastic energy and fracture energies, allowing to calculate $\phi$ using $\psi_o$ as driving force. This equation might depend on the local values of
$\mathbb{C}$, $G_c$ and $\ell$ and, therefore, the model is able to naturally reproduce crack deflection result of the phase distribution within the microstructure \cite{hossain2014effective,na2017effects} also allowing to consider the presence of existing initial cracks \cite{singh2016fracture,zambrano2023arc,aranda2025crack} or emulate Neumann free boundary conditions for $\phi$ between phases using heterogeneous values of $\ell$ \cite{MAGRI2021113759}. The system is completed with Eq.\eqref{eq:SystemDis0c}, where the power of external loads is controlled by a time dependent function, allowing to set on the stress, strain or other type of control \cite{aranda2025crack}. Focusing on homogenization problems, the domain corresponds to a periodic RVE representing a point at the macroscale as depicted in Fig.\ref{fig:PFFscheme}. In this case the external power, following Hill-Mandel conditions, is controlled by the macroscopic strain $\boldsymbol{E}_{M}=\int_{\Omega}\boldsymbol{\varepsilon}d\Omega$ and the microscopic strain fields can then be split in two terms
\begin{equation}
\boldsymbol{\varepsilon}(\vec{x}) = \tilde{\boldsymbol{\varepsilon}}(\vec{x}) + \boldsymbol{E}_{M} 
\label{eq:split}
\end{equation}
being $\tilde{\boldsymbol{\varepsilon}}$ the strain fluctuation field, which average in the RVE vanishes. 

A classical PFF-FFT formulation in homogenization for full strain control \cite{chen2019fft, aranda2025crack} can be defined by monotonically increasing the macroscopic strain. The limitation of this type of control is that it leads to numerical problems and loss of information in cases where the increase of the strain leads to unstable crack growth
. Following Griffith theory for a crack with given size $a$, the condition for crack propagation is given in \ref{eq:griffitha} with $E_{mec}$ being the elastic energy of the RVE defined in Eq.\eqref{eq:elas_Ener}. Crack propagation will continue in a stable manner if the condition on \ref{eq:griffithb} is fulfilled,
\begin{subequations}
\label{eq:griffith}
\begin{eqnarray}
\left. \frac{-\delta E_{mec}}{ \delta a}\right|_\varepsilon = G_c \label{eq:griffitha}\\
\left. \frac{\delta}{\delta a} \left(\frac{-\delta E_{mec}}{ \delta a}\right|_\varepsilon \right)\leq 0, \label{eq:griffithb}
\end{eqnarray}
\end{subequations}
\noindent being the crack surface $a$ represented in PFF model by the definition of $\Gamma$ in Eq.\eqref{eq:CrackLenght}. If the condition in Eq.\eqref{eq:griffithb} is not fulfilled for a certain $E_{M}$, a reduction in its value should be needed to maintain the fracture condition in Eq.\eqref{eq:griffitha}. This situation makes it difficult to follow the fracture process with a monotonic increasing strain control $E_M$. To overpass this problem, some works such as \cite{vignollet2014phase, bharali2022robust}  prescribe the value of the functional $D$ in Eq.\eqref{eq:CrackLenght} as a function of time and solve for the damage and strain variables including the macroscopic strain $\mathbf{E}_{M}$ as an unknown. To reduce the loading state to a single parameter, in \cite{aranda2025crack} the macroscopic strain is set as \begin{equation}\label{eq:f_defintion}
\boldsymbol{E}_M=\boldsymbol{f}E
\end{equation}
where the $\boldsymbol{f}$ tensor allows defining the proportionality between the components of the mean strain tensor and the value $E$ is searched to fulfill the prescribed value of the dissipation functional. Adding this additional variable to the problem, the system is compensated by replacing Eq.\eqref{eq:SystemDis0c} with only one scalar constraint equation. The final governing equations are provided in Eq.\eqref{eq:SystemDis}
\begin{subequations}
\label{eq:SystemDis}
\begin{eqnarray}
\nabla \cdot \boldsymbol{\sigma} =\nabla \cdot \left[g(\phi)\mathbb{C}_{o}:\widetilde{\boldsymbol{\varepsilon}} + g(\phi)\mathbb{C}_{o}:\boldsymbol{f}E \right] = 0 \label{eq:SystemDisa}\\
g'\psi_o+\frac{G_{c}}{\ell}\phi-  \nabla\cdot\left [\ell G_{c}\nabla\phi \right] = 0  \label{eq:SystemDisb}\\
\int_{\Omega} G_{c} \gamma_{f} d\Omega = \tau(t).\label{eq:SystemDisc}
\end{eqnarray}
\end{subequations}
Note that equilibrium equation depends now on strain fluctuation Eq.\eqref{eq:split}. To prevent negative local dissipation (healing) in Eq.\eqref{eq:SystemDisb} the strain energy field $\psi_o$ should be replaced by a history function $\mathcal{H}$, defined following \cite{MieheIJNME2010} as
\begin{equation}
\label{eq:history}
\mathcal{H}(t) = max\left(\hspace{1mm}\psi_o,\mathcal{H}(t_a)\hspace{1mm}\right) \hspace{2mm};\hspace{2mm} t_a\in\left[0,t\right[,
\end{equation}
\noindent where $\mathcal{H}(t_a)$ stands for the history value and $\psi_o$ correspond to the energy in the current time. The treatment of the system of Eq.\eqref{eq:SystemDis} in the FFT numerical scheme is discussed in section \ref{ssec:FFT}.

\subsection{Phase Field Fracture in anisotropic materials}\label{sec:aniPFF} 
Heterogeneity in the material response at the microscale can also arise due to the presence of an anisotropic material with different local orientations, as it happens for example in polycrystalline materials. To this aim, 
the functionals defining the elastic and fracture energies can be modified to include anisotropic elastic energy and anisotropic fracture energy densities.

Focusing on the elastic energy density, the definition of the elastic stiffness in Eq.\eqref{eq:elas_Ener} can be directly redefined in this work to represent a material with cubic anisotropy,  
\begin{equation}
\label{eq:elas_Ener_rot}
E_{mec} = \int_{\Omega} g(\phi)\psi_{o}(\boldsymbol{\varepsilon}) \mathrm{d}\Omega = \int_{\Omega} g(\phi) \frac{1}{2} \mathbb{C}^{r}_{o}:\boldsymbol{\varepsilon}:\boldsymbol{\varepsilon}  \mathrm{d}\Omega,
\end{equation}
\noindent where $\mathbb{C}^{r}$ is the stiffness tensor after a rotation to its actual orientation. This rotation is represented by the tensor $\mathbf{R}$, and the resulting stiffness in global axis is given by Eq.\eqref{eq:Crot}: 
\begin{equation}
\label{eq:Crot}
\frac{\partial^{2} \psi}{\partial \boldsymbol{\varepsilon}^{2}} = g(\phi)\mathbb{C}^{r}_{0}=g(\phi)R_{im}R_{jn}R_{kp}R_{lq}C_{mnpq_{o}}\hspace{1mm}
\end{equation}
\noindent where $\mathbb{C}_{o}$ is the pristine stiffness tensor in lattice axis.

In addition to the mechanical response, anisotropy can appear in the driving force for crack propagation due to the presence of brittle crystallographic planes. In this case, the dissipation functional in Eq.\eqref{eq:SystemDisb} will be modified. These ideas have been applied in \cite{clayton2015phase} for crack orientation in single directional planes, in \cite{Teichtmeister2017} to define systems of crack orientations in orthotropic, cubic and transverse isotropy cases,  and in \cite{MA2020, SeonHong2018} to consider preferential crack propagation planes in multi-PFF models. These models are conceived to represent fracture processes similar to cleavage and aim to generate weakening in the damage field at specific orientations by defining a second-order anisotropy tensor $\mathbf{A}_f$. Based on this, the dissipation density can be modified to emulate PFF-cleavage directions as proposed in \cite{clayton2015phase} and later formally implemented in a PFF model in \cite{SeonHong2018,MA2020}:
\begin{equation}
\label{eq:CrackLenght_ani}
D(\phi,G_{c},\boldsymbol{A}_f) = \int_{\Omega} \gamma_{f}(\phi,G_{c},\boldsymbol{A}_f) d\Omega = \int_{\Omega} \frac{G_{c}}{2\ell}\phi^{2}+\frac{G_{c}\ell}{2} \boldsymbol{A}_f: \left( \nabla\phi\otimes\nabla\phi \right) d\Omega,
\end{equation}
\noindent where we assume a potential case in which $G_c$ could be heterogeneous if this were the case. The tensor $\boldsymbol{A}_f$ is defined using the vector $\vec{n}$ which corresponds to the normal of a weak plane, typically a cubic direction for cubic symmetry. In the case of 2D domains, the projection of the crack vector onto the domain plane $\vec{n}_{pl}$ is used and the anisotropy tensor is defined as
\begin{equation}
\label{eq:PFFani}
\mathbf{A}_f = \boldsymbol{I} + \beta (\boldsymbol{I}-\vec{n}_{pl}\otimes\vec{n}_{pl})
\end{equation}
where $\boldsymbol{I}$ is the second order identity and $\beta$ a weight parameter that allows to set the level of anisotropy in the crack. Damage isotropy can be recovered making $\beta=0$. With these definitions, a new system of governing equations can be defined including modifications to the energy functionals:
\begin{subequations}
\label{eq:SystemDisany}
\begin{eqnarray}
\nabla \cdot \boldsymbol{\sigma} =\nabla \cdot \left[g(\phi)\mathbb{C}^{r}_{o}:\widetilde{\boldsymbol{\varepsilon}} + g(\phi)\mathbb{C}^{r}_{o}:\boldsymbol{f}E \right] = 0 \label{eq:SystemDisanya}\\
g'\mathcal{H}+\gamma_{f}'(\phi,G_{c},\mathbf{A}_f) = 0 \label{eq:SystemDisanyb}\\
\int_{\Omega} \gamma_{f}(\phi,G_{c},\mathbf{A}_f) d\Omega = \tau(t).\label{eq:SystemDisanyc}
\end{eqnarray}
\end{subequations}
\noindent where the derivatives respect to $\phi$ of the dissipation density is
\begin{equation}
\label{eq:gamFunct}
\gamma_{f}'(\phi,G_{c},\mathbf{A}_f) = \frac{G_{c}}{\ell}\phi - \nabla\cdot \left(G_{c}\ell \boldsymbol{A}_f\cdot\nabla\phi \right).
\end{equation}
Since the energy driving force of Eq.\eqref{eq:elas_Ener_rot} is a scalar field, the same definition over the history parameter $\mathcal{H}$ defined in Eq.\eqref{eq:history}, can be used in Eq.\eqref{eq:SystemDisanyb}.

\section{Effective toughness from micromechanical simulations}
\label{ssec:effective_Gc}
As mentioned in the introduction, there are different definitions of the effective toughness of a heterogeneous material. In the approach proposed by Hossain et. al. \cite{hossain2014effective} the J-integral is used on a RVE domain in which the microstructure is explicitly modeled and special boundary conditions are introduced which represent the displacement field of the mode I surface separation (developed in \cite{zehnder2012fracture})  corresponding to the macro-crack advance in a particular direction oriented with the RVE. In this case, the toughness estimate can be obtained by means of a J-integral surrounding this domain, which is considered a measure of the energy release rate $G$ on the macroscopic scale that directly captures the influence of the microstructure. This influence depends directly on the propagation of a microcrack that can be arrested or deflected by differences of stiffness and local toughness. Such microcrack propagation is modeled with a PFF model and its changes in direction and phase crossing can be perceived by the J-integral. In that work, the maximum value over time $t$ of the J-integral is defined as the effective toughness estimate of the microstructure, arguing that the maximum $G$ is a rate of energy that is necessary to overpass at some point to fracture the microstructure. 
\begin{equation}
\label{eq:HosseinGc}
G_{CeH} = \max_{t} J(t).
\end{equation}
This metric considers the influence of the microstructure both due to heterogeneity in stiffness and toughness. However, the use of path integrals is not clear when the microstructure is heterogeneous, which has the disadvantage of having to resort to special materials at the boundary to perform the integral. 

Another alternative for effective toughness definition is the one developed by Schneider \cite{schneider2020fft} with a variational basis, where it is stated that the fracture dissipation functional is independent of the mechanical microfields, idea postulated in \cite{braides1996homogenization}.
The independence of the functional from local strain or stress fields allows to state a minimization problem to find a crack path that represents the minimum fracture energy over microstructures with heterogeneous local toughness fields.  As a result, non-planar crack paths corresponding to a minimal energy state are obtained, and the effective toughness is directly linked with the energy spent on that particular crack path. The definition of this effective energy release rate corresponds to Eq.\eqref{eq:SchneiderGc}, 
\begin{equation}
\label{eq:SchneiderGc}
G_{CeS}(\vec{t_c}) = \lim_{L \rightarrow \infty}  \inf_{S_\phi \subseteq Q_L} \frac{1}{\left|Q_L \right|} \int_{Q_L} G_c \left|\left| \vec{t_c}+\nabla{\Phi} \right|\right| d\Omega.
\end{equation}
 Here $G_c$ is the local toughness, $\Phi$ is a smooth scalar function characterizing the crack (similar to $\phi$ in PFF model) and $Q_L$ is the RVE volume with size $L$ in which the minimization is performed. The fact that this measure includes the limit of $L\rightarrow \infty$ indicates that an adequate measure of effective toughness must be made on a scale much larger than the microstructure. 

The effective toughness estimate considered in this work was sketched first in \cite{aranda2025crack} and takes elements from both approaches. On one hand, the effective toughness will be proportional to the total amount of energy necessary to fracture a particular microstructure (computed with Eq.\eqref{eq:CrackLenght} similar to \cite{schneider2020fft}), but on the other hand, the full mechanical problem in the RVE will be considered in order to capture the effect of heterogeneous local stiffness (as in \cite{hossain2014effective}). From Eq.\eqref{eq:Free_Ener}, the dissipated energy can be expressed,
\begin{equation}
\label{eq:eneraqui}
 D = \Pi_{M} -E_{mec}(\boldsymbol{\varepsilon},\phi).
\end{equation}
\noindent
This energy is then compared with a referential spatial value, as with $|Q_L|$ in Eq.\eqref{eq:SchneiderGc}. Following the idea that a heterogeneous RVE must be replaced with a representative effective toughness $G_{Ceff}$, this value can be associated with the crack path in a homogeneous medium, in which the crack surface and crack direction depend only on the prescribed simulation control conditions in the tensor $\boldsymbol{f}$, Eq. \eqref{eq:f_defintion}. Let the area of the referential crack be $A_{g}$ (a length $L_g$ in 2D cases), its dissipated energy corresponds to $G_{Ceff}(\boldsymbol{f})A_{g}$ and has to be equal to the total energy expended in the actual process $D$. Therefore,  the metric proposed here for the effective fracture toughness is
\begin{equation}
\label{eq:JaviGc0}
G_{Ceff}(\boldsymbol{f}) =\frac{ D}{A_{g}}.
\end{equation}
The fact that $G_{Ceff}$ depends on $\boldsymbol{f}$ highlights the dependency of energy dissipation with loading direction, similar to the dependency of the toughness with the macroscopic crack direction $\vec{t}_c$ in \cite{schneider2020fft}.  

In a pure variational approach, this value can be obtained by evaluating directly the fracture energy in Eqs. \eqref{eq:CrackLenght} or \eqref{eq:CrackLenght_ani}. However, in most cases, the PFF approach introduces some non-variational elements such as the history term defined in Eq. \eqref{eq:history}, the use of hybrid approaches \cite{delorenzis15}, etc., in which these expressions do not coincide exactly with the dissipation. In a totally general case, the dissipated energy at the final fracture stage can be rigorously evaluated using Eq. \eqref{eq:eneraqui}, if the external energy is adequately obtained. This is precisely the advantage of the control method proposed in \cite{aranda2025crack} where this external energy can be obtained as the integration of the external power over a process that is always in equilibrium. The valuation of Eq. \eqref{eq:eneraqui}  will coincide with the dissipation functional in pure variational PFF, but will include other dissipation sources in general cases. Moreover, even for a variational PFF model, under unstable propagation, the uniqueness of the crack path obtained using a staggered scheme is not guaranteed. Therefore, also in this case, the integration of Eq. \eqref{eq:eneraqui} over a stable load path using \cite{aranda2025crack} provides a clearer result from the microscopic viewpoint.

Finally, is important to note that the estimate proposed introduces a dependence on the initial crack. It has been observed that the influence of its size is negligible, since this value drives the stress at which the crack propagates, but not the dissipation. On the contrary, in some cases, there is a non-negligible effect on the crack position which controls the initiation of the crack and therefore introduces a bias in the optimal crack path. To overcome this effect, sampling on different crack positions could be done for the same RVE, defining the estimate as the minimum toughness. 

\section{Numerical framework}\label{sec:NumFrw}

\subsection{FFT based implementation of Phase field fracture and dissipation control}\label{ssec:FFT}
In this section, the FFT implementation of the PFF model with the crack-length control method will be described. The resulting equations and algorithms are an extension of the ones proposed in \cite{aranda2025crack} to consider the general case of a heterogeneous material which could present heterogeneity in stiffness and toughness, as well as anisotropic phases in both elastic response and fracture energy (cleavage).

The objective of the scheme is to find, at each time step ---denoted by the superscript $n$--- the values of $\widetilde{\boldsymbol{\varepsilon }}$, $\phi$ and $E$ which fulfill Eqs. \eqref{eq:SystemDisanya} and \eqref{eq:SystemDisanyb} for a given value of the macroscopic fracture energy $D^n$ (Eq.\eqref{eq:CrackLenght} and Eq.\eqref{eq:CrackLenght_ani}).  The procedure is equivalent to an arc-length, as proposed in \cite{singh2016fracture} for PFF-FEM. The non-linear system of Eq.\eqref{eq:SystemDisany} can be solved using a Newton-Raphson scheme (NR) implemented on an FFT-based algorithm. This is achieved by defining a non-linear differential operator for the NR residual $R(q)$ from the system in Eq.\eqref{eq:SystemDisany}, which depends on the unknown set of variables $q=\{\widetilde{\boldsymbol{\varepsilon}},\phi,E\}$. Then, for each Newton iteration $i$, the linearization of this residual around the current value of the set $q_i$ provides a Jacobian bi-linear form $J(q_i,\Delta q)$ which depends on the variation of the unknown variable set $\Delta q=\{\Delta \widetilde{\boldsymbol{\varepsilon}},\Delta \phi,\Delta E\}$. Both operators are formulated using the Fourier space. The linearized expression of the residual is then set to 0, resulting in a linear system
\begin{equation}
\label{eq:schemeSys}
R = 0  \hspace{2mm}\Rightarrow\hspace{2mm} R(q_i+\delta q)\approx R(q_{i})+\left[ J (q_{i},\Delta q) \right]=0,
\end{equation} 
\noindent where the increment $\Delta q$ is the unknown to be found at each NR iteration. To obtain the Jacobian, each equation in the system Eq.\eqref{eq:SystemDisany} has to be linearized. The first equation (Eq.\eqref{eq:SystemDisanya}) represents the mechanical equilibrium and is formulated with the Fourier-Galerkin method \cite{VONDREJC2014,ZemanGeus2017}. In this method, the equilibrium is written as a convolution of the stress with a projection operator $\mathbb{G}*\sigma$. The stress to enter in Fourier Galerkin is the one linearized with respect to the current strain and damage values as
\begin{equation}
\label{eq:linSig}
\boldsymbol{\sigma}_{i+1} \approx \boldsymbol{\sigma}_{i} + g(\phi)\mathbb{C}^{r}_{oi}:\Delta\boldsymbol{\varepsilon}_{i+1} + g'(\phi_{i})\boldsymbol{\sigma}_{oi}\Delta\phi_{i+1}  \hspace{3mm};\hspace{3mm} \Delta\boldsymbol{\varepsilon}_{i+1}=\Delta\widetilde{\boldsymbol{\varepsilon}}_{i+1}+\boldsymbol{f}\Delta E_{i+1}, 
\end{equation}
\noindent where $\mathbb{C}^{r}_{oi}$ is the stiffness tensor of the pristine material in the global axis of the $i^{th}$ NR iteration. Eq.\eqref{eq:SystemDisanyb} can be used to define the residual and Jacobian of the damage related parts of the system and Eq.\eqref{eq:SystemDisanyc} to define the residual and Jacobian involving the fulfillment of the dissipation control.
The residual and Jacobian will be computed in real space, but using the Fourier space in order to solve numerically the divergence, gradient and convolution operations present in the system. With these considerations and using the expressions in Eq.\eqref{eq:linSig}, the residual for the heterogeneous/anisotropic materials formulation reads
\begin{gather}
\label{eq:aniSystem_CrackLength}
R_{1}(q_{i}) = \mathcal{F}^{-1} \left\{ \widehat{\mathbb{G}} : \mathcal{F}\left[ \mathbb{C}^{r}_{i}:\widetilde{\boldsymbol{\varepsilon}}_{i}+\mathbb{C}^{r}_{i}:\boldsymbol{f}E_{i} \right] \right\} \nonumber \\
R_{2}(q_{i}) = g'(\phi_{i})\mathcal{H}_{i} + \frac{G_{c}}{\ell}\phi_{i} - \mathcal{F}^{-1}\left[\vec{\xi}\cdot\mathcal{F}\left[G_{c}\ell \boldsymbol{A}\cdot \mathcal{F}^{-1}\left[\vec{\xi}\widehat{\phi}_{i} \right]  \right] \right] \nonumber \\
R_{3}(q_{i}) = \int_{\Omega} G_{c} \left[    \frac{1}{2\ell}\phi^{2}_{i}+ \frac{\ell}{2} \boldsymbol{A}:\left( \mathcal{F}^{-1}\left[\vec{\xi}\hat{\phi}_{i}\right] \otimes \mathcal{F}^{-1}\left[\vec{\xi}\hat{\phi_{i}}\right] \right) \right] d\Omega - \tau^n,
\end{gather}
\noindent being the symbols $\mathcal{F}$, $\mathcal{F}^{-1}$ the Fourier and inverse Fourier operations, $\vec{\xi}$ is the complex frequency vector and $R_{1}$, $R_{2}$ and $R_{3}$ respectively to the residual of equilibrium, residual of damage equation and the control equation residual. Similarly, the Jacobian is defined in Eq.\eqref{eq:aniCrackLength_linearization} as
\small
\begin{gather}
 J_{1}(q_{i},\Delta q_{i+1}) = \mathcal{F}^{-1} \left\{ \widehat{\mathbb{G}} : \mathcal{F}\left[ \mathbb{C}^{r}_{i}:\Delta\boldsymbol{\varepsilon}_{i+1}+g'(\phi_{i})\frac{\partial \psi_{oi}}{\partial \boldsymbol{\varepsilon}_{i}} \Delta\phi_{i+1} \right]  \right\} \nonumber \\ 
 J_{2}(q_{i},\Delta q_{i+1}) = g'(\phi_{i})\frac{\partial \mathcal{H}_{i}}{\partial \boldsymbol{\varepsilon}_{i}}:\Delta\boldsymbol{\varepsilon}_{i+1}+ \left( \frac{G_{c}}{\ell} + 2\mathcal{H}_{i} \right) \Delta\phi_{i+1} - \mathcal{F}^{-1}\left[\vec{\xi}\cdot\mathcal{F}\left[G_{c}\ell \boldsymbol{A}\cdot \mathcal{F}^{-1}\left[\vec{\xi}\hspace{1mm}\widehat{\Delta\phi}_{i+1} \right]  \right] \right]  \nonumber \\
 J_{3}(q_{i},\Delta q_{i+1}) = \int_{\Omega} \frac{G_{c}}{\ell}\phi_{i}\Delta\phi_{i+1} - G_{c}\ell \mathcal{F}^{-1}\left[\vec{\xi}\cdot\mathcal{F}\left[ \boldsymbol{A}\cdot \mathcal{F}^{-1}\left[\vec{\xi}\widehat{\phi}_{i} \right]  \right] \right]  \Delta\phi_{i+1} d\Omega,
 \label{eq:aniCrackLength_linearization}
\end{gather}
\normalsize
\noindent where $\Delta\boldsymbol{\varepsilon}_{i+1}=\Delta\widetilde{\boldsymbol{\varepsilon}}_{i+1}+\boldsymbol{f}\Delta E_{i+1}$ and the expression of the projection operator $\widehat{\mathbb{G}}$ in Fourier space is given by Eq.\eqref{eq:GFourier}. 
\begin{equation}
\label{eq:GFourier}
\widehat{\mathbb{G}} = 0 \hspace{2mm}\mathrm{for}\hspace{2mm}\vec{\xi}= 0 \hspace{2mm};\hspace{2mm}\widehat{\mathbb{G}}_{ijkl}=\delta_{ik}\frac{\xi_{j}\xi_{l}}{\vec{\xi}\cdot\vec{\xi}}\hspace{2mm}\mathrm{for}\hspace{1mm}\vec{\xi}\neq \vec{0}.
\end{equation}
Note that in all of these definitions the energy $\psi_o$ has been replaced by the history variable $\mathcal{H}_{i}$ for the damage related parts of the operators. In the Jacobian term $J_{2}$ the derivative of the history with respect the strain field has also been included,  fundamental to achieve convergence. The actual expressions of the history in the NR scheme and the definition of its derivative correspond to,
\begin{equation}
\mathcal{H}_{i} = max(\psi_{oi},\mathcal{H}^{n}) \hspace{2mm};\hspace{2mm}
\frac{\partial \mathcal{H}_{i}}{\partial \boldsymbol{\varepsilon}_i}=
    \begin{cases}
        0 & \text{if } \psi_{oi} \leq \mathcal{H}_{i}\\
        \boldsymbol{\sigma}_{oi} & \text{if } \psi_{oi} > \mathcal{H}_{i}.
    \end{cases}
\label{eq:Delta_history}
\end{equation}
 
 At each time step and NR iteration, the linear system with $7(n_x \cdot n_y \cdot n_z)+1$ unknowns given in Eq.\eqref{eq:schemeSys} is solved to find $\Delta q$. Since this system is non-symmetric, the bi-conjugate gradient stabilized method is used \cite{aranda2025crack}. Then, the new value of the unknown set, $q_{i+1}$ is determined by adding the increment $\Delta q_{i+1}$ to the current $q_{i}$, leading to  
\begin{equation}
\label{eq:sumq}
q_{i+1} =q_{i}+\Delta q ; \ q_0=q^n,
\end{equation}
\noindent where $q^{n}$ corresponds to the converged value  on the previous time step. After computing the new $q_{i+1}$, the fields depending on $q$ are also updated  using equations in section \ref{sec:MODELS},
\begin{equation}
\label{eq:upNq}
\boldsymbol{\sigma}_{i+1}=\boldsymbol{\sigma}(q_{i+1}) \hspace{2mm};\hspace{2mm}
\mathbb{C}_{i+1}=\mathbb{C}(q_{i+1}) \hspace{2mm};\hspace{2mm}
\psi_{oi+1}=\psi_{o}(q_{i+1}) \hspace{2mm};\hspace{2mm}
\mathcal{H}_{i+1} = max(\psi_{oi+1},\mathcal{H}^{n}),
\end{equation}

\noindent where $\mathcal{H}^{n}$ is the history of the last converged time. The criteria to stop iterating in the NR scheme is defined with the error in Eq.\eqref{eq:errors_crack} with respect to a defined tolerance:
\begin{equation}
\label{eq:errors_crack}
err  = max \left(  \frac{||\Delta\boldsymbol{\varepsilon}_{i+1}||}{||\boldsymbol{\varepsilon}_{i+1}||}  , \frac{||\Delta\phi_{i+1}||}{||\phi_{i+1}||} , \frac{D_{i+1}-\tau^n}{\tau^{n}}\right)<tol.
\end{equation}
If the Newton iterative process converges, the resulting set of unknowns is updated in time as $q^{n+1}=q_{i+1}$ together with the history as $\mathcal{H}^{n+1}=\mathcal{H}_{i+1}$.

The convergence of the Newton-Raphson procedure can be improved in some cases using a successive relaxation strategy, in which a parameter $\alpha$ is used to dampen the solution by the penalization of each obtained $\Delta q$ in NR iterations. To this aim an adaptive under-relaxation strategy is implemented, similar to the one developed in \cite{bharali2022robust, storvik2021accelerated} for staggered PFF models, is implemented here. The parameter $\alpha$ is defined at each time step as a function of the maximum macroscopic stress in previous time
\begin{equation}
\label{eq:relax}
\alpha^n  = max\left(\theta,\frac{\boldsymbol{\Sigma}_{max}}{\boldsymbol{\Sigma}^{n-1}}\right) \hspace{2mm};\hspace{2mm} \boldsymbol{\Sigma}_{max}=max([0,\boldsymbol{\Sigma}^{n-1}]),
\end{equation}
\noindent where $\theta$ is the minimum relaxation parameter allowed.
With this definition,  the value of $\alpha_n$ will be $1$ until propagation occurs, and then it will decrease as it approaches coalescence. For more details, the reader is referred to the work  \cite{aranda2025crack}.

\subsection{Computation of effective toughness}\label{sec:computTough}
\begin{figure}[htbp]
\centering
\includegraphics[width=110mm]{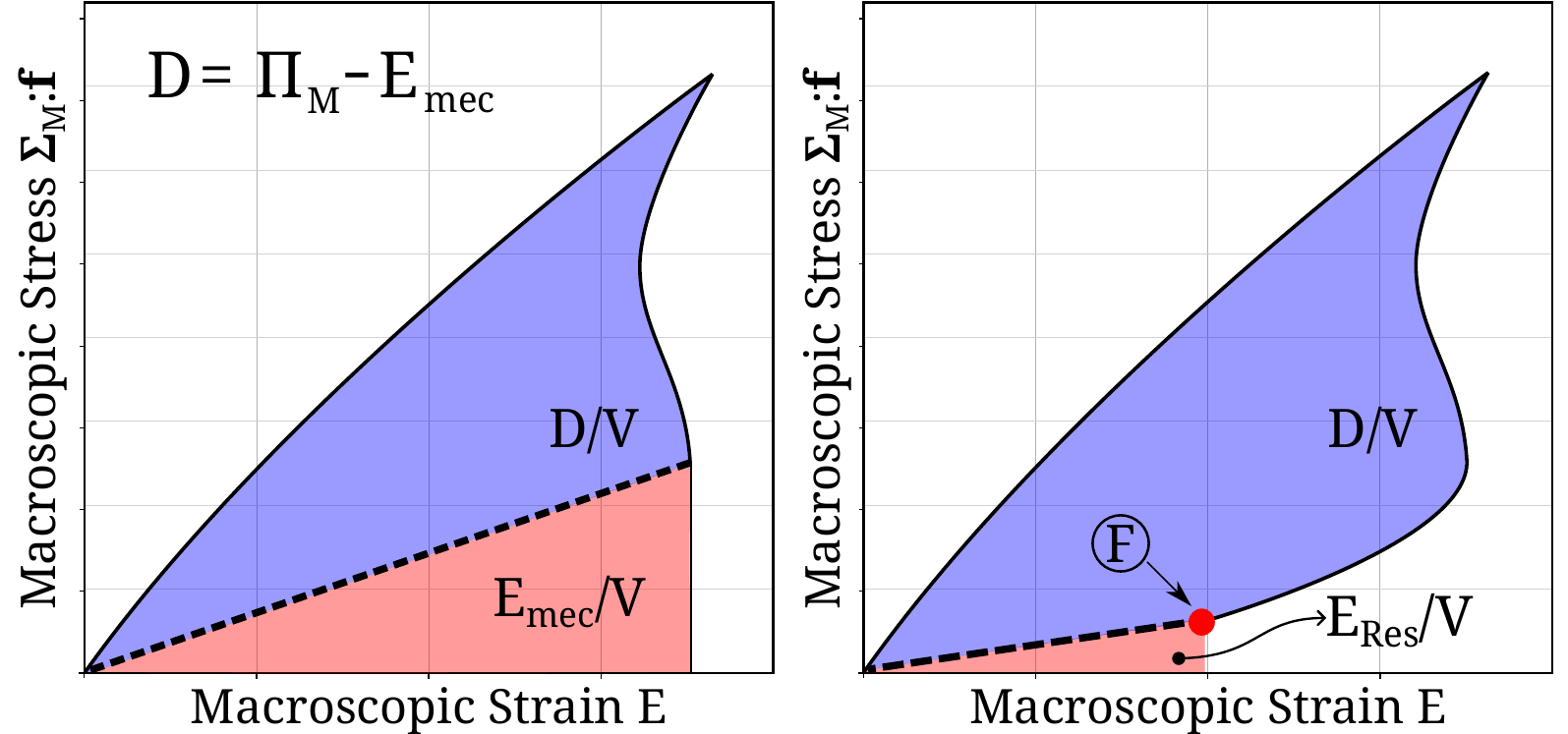}
\caption{\centering{External power $\Pi_M$, stored energy $E_{mec}$ and dissipation $D$ represented by areas under the curve of a snap-back/fracture behavior of an RVE of volume $V$.}}
\label{fig:EffTougnesScheme}
\end{figure}

The effective toughness estimate proposed, Eq. \eqref{eq:JaviGc0} requires the evaluation of the work of external stresses over the RVE, $\Pi_{M}$, in Eq. \eqref{eq:eneraqui}. This energy can be computed under an equilibrium loading path by the integration of the internal power. The external energy introduced in the full RVE can be obtained from the integration of the internal power based on the macroscopic $\boldsymbol{\Sigma}_M$ and $\boldsymbol{E}_M$, as depicted in Fig.\ref{fig:EffTougnesScheme}. In this figure, the lower area is related to the elastic energy $E_{mec}$ stored at a given point of the fracture process, Eq.\eqref{eq:elas_Ener}. By thermodynamical equilibrium, the upper part of the area must correspond to the energy density of the dissipative processes, which for elastic materials should only be attributed to fracture dissipation also represented by Eq.\eqref{eq:CrackLenght}. However, as previously stated, the introduction of a local history might influence this statement as it will be analyzed later.

Using the PFF-FFT model with dissipation control allows us to obtain the exact fracture point of the full RVE, providing the energy which corresponds to the complete fracture process. Beyond that point, the (small) elastic energy remaining in Eq.\eqref{eq:elas_Ener} becomes the value $E_{res}$ (Fig. 2), attributed to the residual stiffness of the cracked area. The fracture point, marked as $F$ in Fig.\ref{fig:EffTougnesScheme}, is reported as the moment when the coalescence of PFF cracks occurs either with the domain boundary \cite{zambrano2023arc} or with other PFF cracks \cite{aranda2025crack} and usually corresponds to a strain lower than the maximum, typical in cases with unstable crack growth, so it is not always possible to find $F$ in a controlled strain/displacement scheme \cite{singh2016fracture, aranda2025crack}. 

The definition of the effective toughness is made according to Eq. \eqref{eq:JaviGc0} using the integration of the external power,
\begin{equation}
\label{eq:JaviGc}
G_{Ceff}(\boldsymbol{f}) = 
\frac{V}{A_{g}} \left(\int_{E=0}^{E=E_f} \boldsymbol{\Sigma}_M:\boldsymbol{f} dE  - E_{res}\right),
\end{equation}
\noindent where $V$ is the volume of the RVE, $E_{f}$ is the value of the strain $E$ at the crack coalescence and $A_{g}$ the referential crack surface at Eq. \eqref{eq:JaviGc0}. Note that $A_{g}$ refers to the surface generated during the fracture process associated with $D$ once fracture ends (Fig. \ref{fig:EffTougnesScheme}), and must exclude any pre-existing cracks. In this expression, the value of the  has to be numerically inferred from the calculated macroscopic strain-stress curve. To this aim, Simpson's rule is used to compute the area under the curve until the fracture point, from which the residual energy must be removed:
\begin{equation}
\label{eq:ABC}
\int_{0}^{E^{f}} \boldsymbol{\Sigma}_M:\boldsymbol{f} dE - E_{res}= T^{f}- \frac{1}{2}\Sigma^{f}E^{f} \hspace{3mm};\hspace{3mm} T^f= \sum_{j=1}^{f} \frac{1}{2}(\Sigma^{j}+\Sigma^{j-1})(E^{j}-E^{j-1}),
\end{equation}
\noindent where $T$ is the total area under the curve and the super-index $n=f$ in every variable represents a value of the iterative procedure in section \ref{sec:NumFrw} at the step time $n$ in which fracture occurs.

In comparison with the work Hossain et. al. \cite{hossain2014effective} it is important to remark that Eq.\eqref{eq:JaviGc} corresponds to the full process, while in \cite{hossain2014effective} the fracture energy is defined based on the instantaneous values of the energy release rate at every time \eqref{eq:HosseinGc}. In order to compare the results, our model also allows us to obtain estimations of the instantaneous energy release rate based on Eq.\eqref{eq:griffith} by computing the variation of the dissipation with respect to the variation of the total crack surface. This can be calculated with a numerical derivative $\partial T^n/\partial \Gamma $ as 
\begin{equation}
\label{eq:Gmed}
G = \frac{-\delta E_{mec}}{ \delta \Gamma}=\frac{V (T^{n}-T^{n-1})}{\Gamma^{n}-\Gamma^{n-1}}\hspace{3mm};\hspace{3mm} T^n=\sum_{j=1}^{n} \frac{1}{2}(\Sigma^{j}+\Sigma^{j-1})(E^{j}-E^{j-1}),
\end{equation}
where the value of the crack length $a$ has been replaced by its PFF equivalent in Eq.\eqref{eq:CrackLenghtreal} or Eq.\eqref{eq:CrackLenght_ani}. 
This value has an equivalence to the one obtained with the procedure in Eq.\eqref{eq:HosseinGc}, which has already been demonstrated in the case of a homogeneous material in \cite{aranda2025crack}. 

Regarding the definition of the effective toughness in \cite{schneider2020fft}, Eq. \eqref{eq:SchneiderGc}  corresponds to the total energy dissipation weighed on a specific volume, so it should be equivalent to the value of the functionals in Eq.\eqref{eq:CrackLenght} or Eq.\eqref{eq:CrackLenght_ani} over the same volume, which for variational PFF coincides with the area under the curve described above. For this reason, both the method in \cite{schneider2020fft} and our approach should converge to the same value for a given crack path. However, notable differences with our method have also to be highlighted. Since an adequate interpretation of the elastic PFF model based in \cite{miehe2010phase} cannot consider crack nucleation, an initial crack should be included in the RVE, which controls the position of fracture initiation, making it so that the studied PFF crack doesn't correspond to the minimal energy crack in the RVE as in the mentioned work. Another difference is that no macroscopic crack direction is explicitly imposed, but arises as a result of the loading direction dictated by $\mathbf{f}$. This macroscopic strain direction influences the toughness measure consequently with the influence that the macroscopic strain at a point would have on the crack resistance assuming scale separation. Finally, another important difference is that in our approach the heterogeneity of elastic properties at the microscale is considered and might affect the toughness estimated by Eq.\eqref{eq:JaviGc0}. This effect will be analyzed in section \ref{sec:Res}.

\subsection{Modeling pre-existing cracks}\label{ssec:NumConsider}
The PFF model for brittle fracture requires a pre-existing crack to propagate fracture. The most rigorous and efficient method to introduce this crack is still an open issue. Often, notches or sharp contours are used to emulate a crack  \cite{delorenzis15, MieheIJNME2010, zhang2019fracture}, especially in FEM where adaptive meshes can be used. Other strategies consist in using boundary conditions, e.g., unconnected FEM nodes on a symmetry boundary \cite{aranda2025crack}, or to emulate the presence of a crack indirectly with local conditions, e.g. localized crack opening \cite{Borden2018}.  

Alternatively, a notch can also be modeled by introducing regions with negligible stiffness as in \cite{chen2019fft, aranda2025crack}, very convenient for 
 grid methods as FFT solvers. Nevertheless, this procedure has certain limitations. First, the minimum crack thickness corresponds to one voxel, which makes the crack sharpness mesh dependent. Also, the severe phase contrast might induce noise and convergence problems in FFT solvers \cite{MA2020}. 
 
 Any of these approaches imply that no pre-existing damage field is present in the domain and, before a PF crack propagates, a large amount of relatively dispersed damage is generated near the concentrator due to elastic energy accumulation. This effect emulates a nucleation phenomenon and could be understood as the nucleation of a PFF crack. However, there is no clarity in this respect and while some studies claim that nucleation of a PFF crack is possible and well represented \cite{tanne2018crack}, more recent studies claim that this process does not correspond to a real physical process \cite{lopez2025classical}. Additionally, the numerical effects of this disperse 
 damage have been reported in works like \cite{singh2016fracture, zambrano2023arc} showing that this process requires additional energy and alters the crack propagation point. 

For these reasons and in order to be able to provide accurate estimations of toughness using the mechanical response, this issue must be contained. The strategy followed here is the use of Phase Field enrichment strategies in which the domain includes specific distributions of an initial damage field to emulate pre-existing fissures, allowing us to avoid the phenomenon PFF nucleation \cite{singh2016fracture}. Furthermore, since in the model a propagating crack is defined with a PF damage distribution, it is more adequate to model an initial crack similarly. Several works in the literature follow these strategies, highlighting the Crack Tip Enrichment procedure (CTE) proposed \cite{singh2016fracture, zambrano2023arc} in which equations equivalent to Eq.\eqref{eq:SystemDisb} are used to generate disperse damage fields from spatial discrete functions in the crack tip. 
The same strategy can be used to enrich an entire discrete notch as in \cite{delorenzis15, hossain2014effective} for FEM models. Other smoothening equations can be used, e.g. in \cite{borden2012phase} where linear distributions of damage are used to enrich the complete contour of an initial crack. Finally, other studies propose the use of penalization methods, which directly intervene in the damage equation \eqref{eq:SystemDisb} to introduce penalized zones imposing damage as Dirichlet boundary conditions \cite{MA2020}. The last work proves that this implementation is also well suited for FFT solvers, where it was effectively applied to model pores and holes. 

In this work, different crack tip enrichment (CTE) approximations and a complete Phase Field Enrichment (PFE), in which the complete initial crack is embedded in an initial damage field, will be tested in order to determine a minimal condition in our FFT approach in which the PFF nucleation does not affect the measure of effective toughness. The basis of all the approximations tested is the generation of a non-localized continuum damage distribution $\phi_{o}$ from a prescribed discrete damage field $\phi_{disc}$ (representing the sharp crack or its boundary). To this aim, the formulation in \cite{MAGRI2021113759} for obtaining non-local damage fields is used throughout Eq. \eqref{eq:damini}
\begin{equation}
\label{eq:damini}
\phi_{o} - \ell \nabla^{2}\phi_{o} = \phi_{disc}  \hspace{2mm}\rightarrow\hspace{2mm} \phi_{ini}=\phi_{M} \frac{\phi_{o}}{max(\phi_{o})},
\end{equation}

\noindent where $\ell$ is taken equal to the characteristic length in the PFF model. The  continuous distribution resulting from Eq. \eqref{eq:damini}  preserves the average of the sharp field $\phi_{disc}$, therefore having a maximum of $\phi_{o}$ smaller than the maximum of the discrete distribution. For this reason, the resulting distribution is re-scaled, as shown on the right side of Eq.\eqref{eq:damini}, being $\phi_M$ a parameter to control the maximum of the initial damage distribution $\phi_{ini}$.

\begin{figure}
\centering
\includegraphics[width=10cm]{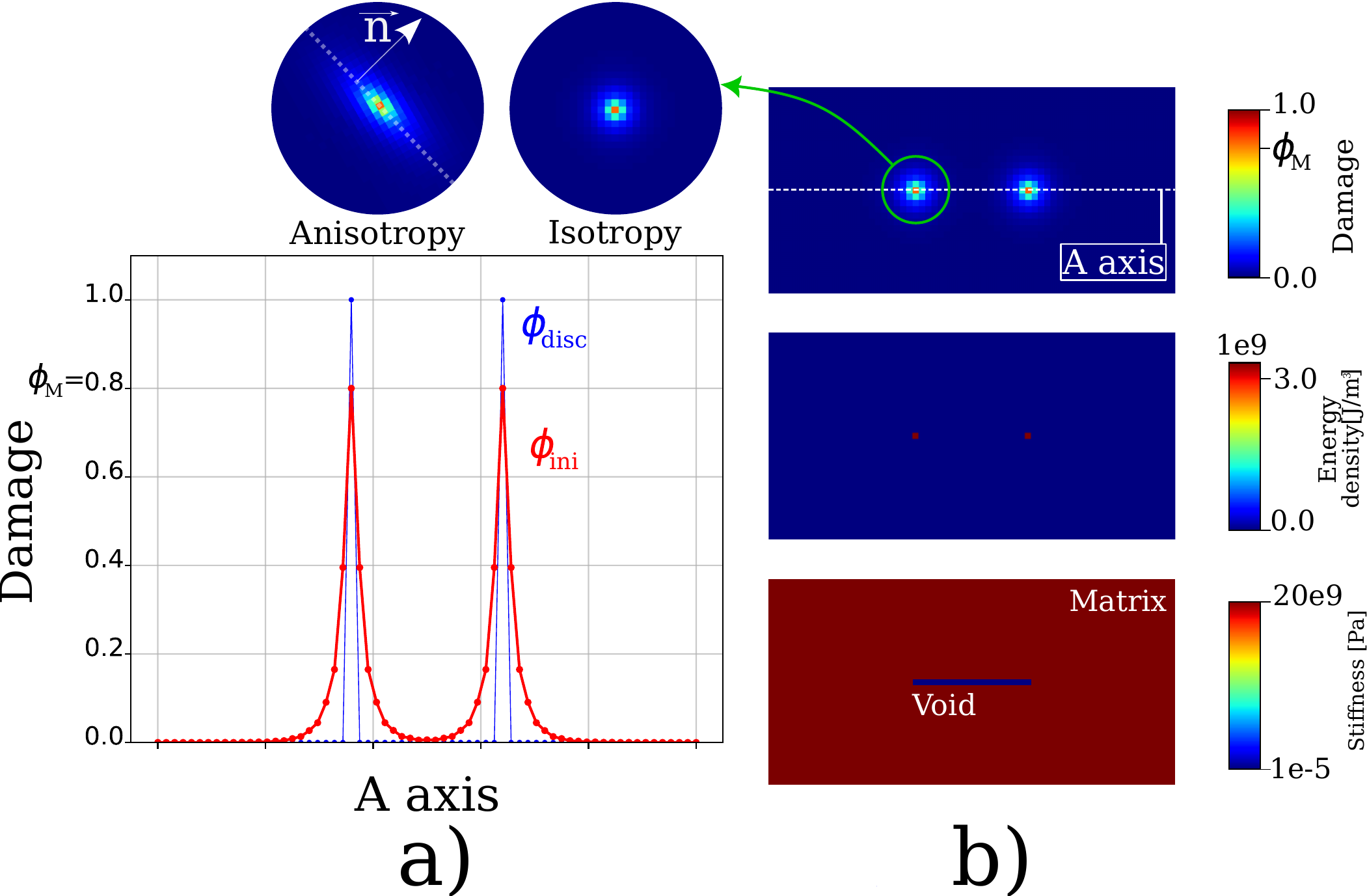}
\caption{\centering{CTE condition for a straight crack. a) Sharp damage field and its corresponding smooth counterpart using Eq. \ref{eq:damini}. b) Representation of the spatial distribution of initial damage fields, initial energy density and phases distributions.}}
\label{fig:damini}
\end{figure}

The smooth damage fields obtained using this process are represented in Fig. \ref{fig:damini}a. In this example, the crack is defined as a row of voxels with negligible stiffness, and the sharp damage distribution, $\phi_{disc}$, is chosen to be zero everywhere with the exception of two points representing the crack tip in 2D,  where $\phi_{disc}=1$. 

The formulation in  Eq.\eqref{eq:damini} will result in an isotropic distribution for $\phi_{ini}$ and does not consider the anisotropic tensor defined for the cleavage formulation in section \ref{sec:aniPFF}. For this reason the modified formulation of this equation used in \cite{teichtmeister2017phase} is made for this kind of materials.
\begin{equation}
\label{eq:daminiAni}
\phi_{o} - \ell \nabla\cdot(\boldsymbol{A}\nabla\phi_{o}) = \phi_{disc}  \hspace{2mm}\rightarrow\hspace{2mm} \phi_{ini}=\phi_{M} \frac{\phi_{o}}{max(\phi_{o})},
\end{equation}

In addition to the definition of the initial damage distribution, the history variable $\mathcal{H}$ from Eq.\eqref{eq:HistoryNumIni} has to be modified accordingly to prevent that the initial damage introduced will disappear during the loading. A similar procedure to the one in \cite{MA2020} or \cite{borden2012phase} is defined here to include $\phi_{ini}$ in the history. The energetic conjugate of $\phi_{ini}$ is defined considering Eq.\eqref{eq:SystemDisb} or Eq.\eqref{eq:SystemDisanyb}, and the initial value of the history ($\mathcal{H}$) is set to that value. This initial energy is inferred from the $\phi_{ini}$ field using Eqs. \eqref{eq:Eninia} or \eqref{eq:Eninib} for a heterogeneous or anisotropic material respectively.
\begin{subequations}
\label{eq:Enini}
\begin{eqnarray}
\psi_{ini} = \frac{G_{c}}{g'(\phi)\ell}\phi -  \frac{\ell}{g'(\phi)}\nabla\cdot\left [ G_{c}\nabla\phi \right] = 0 \label{eq:Eninia}\\
\psi_{ini} = -\frac{1}{g'(\phi)\ell}\gamma_{f}'(\phi,G_{c},\mathbf{A}_f) ,   \label{eq:Eninib}
\end{eqnarray}
\end{subequations}
\noindent where the expression for $\gamma_{f}'$ is the one in Eq.\eqref{eq:gamFunct}. Finally, to include this initial energy distribution to the history $\mathcal{H}$, Eq.\eqref{eq:Delta_history} is modified to the definition in Eq.\eqref{eq:HistoryNumIni},
\begin{equation}
\label{eq:HistoryNumIni}
\mathcal{H}_{i} = max(\psi_{oi}+\psi_{ini}\hspace{1mm},\hspace{1mm}\mathcal{H}^{n}).
\end{equation}
 Including the conjugate energy additively ensures that new values of damage are calculated based on the initial distribution. 

An alternative to the CTE described, which singular points corresponding with the crack tip of the initial crack as in \cite{singh2016fracture}, the full surrounding are of the initial crack can be introduced as initial damage as in \cite{hossain2014effective} defining $\phi_{disc}=1$ in all voxels representing the sharp crack with  negligible stiffness. This procedure is called here Complete Phase Field Enrichment (CPFE) and will be compared with the CTE in section \ref{ssec:IniCrack}.

\section{Results: effective toughness estimations}\label{sec:Res}
In this section the proposed models and measure procedures will be applied in different scenarios. First, the model will be used in homogeneous materials to fit the effective toughness estimates with the imposed PFF toughness as reference and also to test the influence of the initial crack approach considered. Secondly, the method will be applied to 2D laminates in which the effective toughness can be evaluated analytically \cite{schneider2020fft, hossain2014effective}, in order to verify the accuracy of the measure and to test the effect of stiffness and local toughness heterogeneity in the measure. Finally, the procedure developed will be used to obtain effective toughness in heterogeneous fiber composites and anisotropic elastic polycrystals.

All cases of homogeneous material or composites will use the material properties in the table \ref{table:1} as a matrix, being the properties of inclusions changed and specified for each different case.  A unique macroscopic loading condition will be imposed for all cases, which corresponds to uniaxial macroscopic strain and will be represented by 
\begin{equation}
\label{eq:f_feat}
\boldsymbol{E}_{M}=\boldsymbol{f}E=\begin{pmatrix}0&0&0\\0&E&0\\0&0&0
\end{pmatrix}.
\end{equation}

Two kinds of periodic RVEs will be used in 3D to resemble 2D conditions. First kind will be rectangular plates, validated as a less expensive case that is also less prone to unstable fracture \cite{aranda2025crack}, with dimension $L = [L_x,L_y,L_z] = [L_o,L_o/2,L_o/n_x]$, being $L_o=500\mu$m a value fixed in all cases and being $n_x$ the amount of voxels in the X direction to make $L_z$ to be represented by a single voxel. The second geometry will be quadrilateral plates with dimensions $L = [L_o,L_o,L_o/n_x]$. Note that in table \ref{table:1} the value of $\ell$ in the matrix material is defined with a factor $f_f$ and $L_o$, to be dependent on the case discretization that will be specified for each case considered. Regarding the numerical under-relaxation formulation, a minimum relaxation of $\theta=0.3$ is used in every case.

\begin{table}[ht]
\centering
\begin{tabular}{|c|c|c|c|c|} 
\hline
$E_{Young}$ & $\nu$ & $G_{c}$     & $\ell$ \\
$[Pa]$     & $   $ & $[J/m^{2}]$ & $    $ \\
\hline
20e9 & 0.25 & 2e3 & $f_f L_o/n_x$ \\
\hline
\end{tabular}
\caption{Mechanical and PFF properties of matrix material.}
\label{table:1}
\end{table}

\subsection{Validation on the toughness measure}\label{ssec:val}

\subsubsection{Homogeneous material validation and crack tip enrichment analysis}\label{ssec:IniCrack}
\begin{figure}[ht]
\centering
\includegraphics[width=10cm]{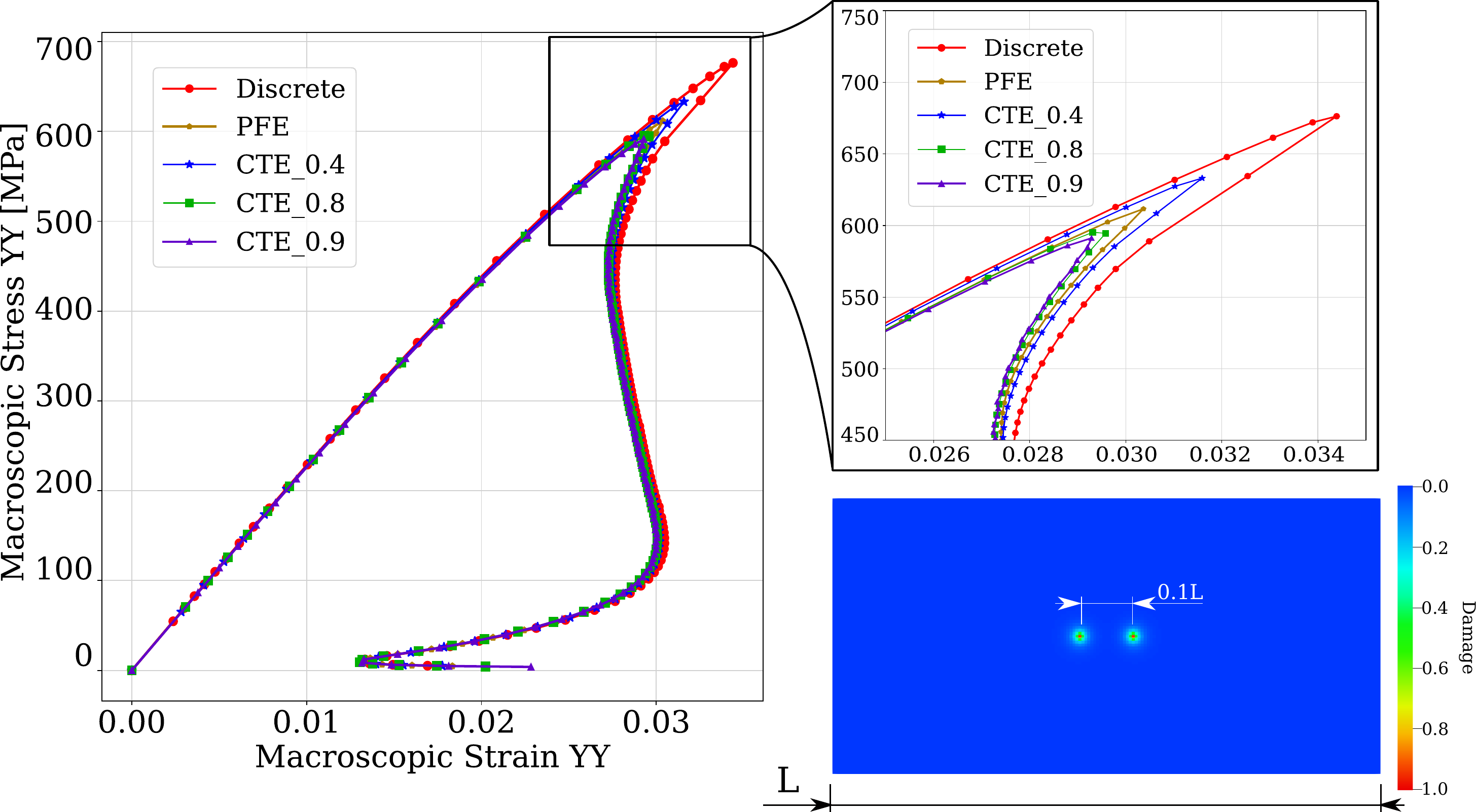}
\caption{\centering{Stress/Strain curves for simulations of initial condition, where each curve is: Simulation with no initial damage(Red), CPFE simulation and $\phi_M=0.98$ (Brown), CTE simulation and $\phi_M=0.4$ (Blue), CTE simulation and $\phi_M=0.8$ (Green), CTE simulation and $\phi_M=0.95$ (Purple).}}
\label{fig:CTEtests}
\end{figure}

The proposed estimate of effective toughness in section \ref{ssec:effective_Gc} will be applied here to homogeneous materials. A rectangular RVE with $185x93x1$ voxels is used, considering only the matrix material defined in Table \ref{table:1}. The characteristic length will be two voxels, $f_f=2$,  similar to the minimum of 2 elements proposed in \cite{miehe2010phase}. The use of the methodology proposed for a case with a homogeneous material should recover the material toughness used in the PFF model. However, it is known that simulation measures actually converge to a slightly different value, which depends on the discretization used in FE from \cite{ bourdin2008variational} or FFT \cite{aranda2025crack}. Moreover, the effective toughness estimates predicted with our approach would also be affected by the \emph{nucleation}, so a set of simulations has been performed considering a panel with different approximations for the initial crack, including a discrete crack of 1 voxel thickness to different smoothed cracks, and the results are represented in Fig.\ref{fig:CTEtests}.

It can be seen that all the curves converge to the same result at the beginning of the loading process and after crack propagation, but the case in which no initial diffuse crack leads to higher stress and strain than others at the crack propagation point. As expected, this implies a larger area under the curve for this case, which also implies a higher perceived crack dissipation. This area is reduced when some initial damage field around the crack is introduced. A full damage around the crack, CPFE, alleviates this problem, but CTE cases with $\phi_M=0.8$ and $\phi_M=0.95$ further improve this result with similar results. Values for the effective toughness estimates in these simulations are depicted in table \ref{table:2} :

\begin{table}[ht]
\centering
\resizebox{\columnwidth}{!}{
\begin{tabular}{|c|c|c|c|c|c|c|}
\cline{2-7} 
\multicolumn{1}{c|}{\multirow{2}{*}{ }}  &  Imposed PFF & Discrete          &     CPTE       &      CTE     &  CTE        &    CTE      \\
\multicolumn{1}{c|}{\multirow{2}{*}{ }} &  (Corrected) & initial condition & $\phi_M=0.2$ & $\phi_M=0.4$ & $\phi_M=0.8$ & $\phi_M=0.95$ \\
\hline
 Toughness [$J/m^2$]&    2296.6         &    2425.74    &    2317.6    &    2353.4    &    2309.9    &    2302.1       \\ 
\hline
 Percent difference &   ---      &    5.65\%         &     0.94\%    &     2.50\%   &    0.61\%    &    0.23\%     \\
\hline
\end{tabular}}
\caption{Effective toughness estimates for homogeneous materials.}
\label{table:2}
\end{table}

It is observed that using CTE with values higher than $\phi_M=0.8$ produce results with errors below $1\%$, being the best option  CTE with $\phi_M=0.95$.

The effect of the characteristic length is then analyzed, exploring the results for $\ell=2L_o/n_x$, $\ell=3L_o/n_x$ and $\ell=4L_o/n_x$ considering a CTE with $\phi_M=0.80$ and $\phi_M=0.95$. The results are shown in Table \ref{table:3}, using as reference the material toughness with the discretization correction in \cite{aranda2025crack} for these parameters.

\begin{table}[ht]
\centering
\resizebox{\columnwidth}{!}{
\begin{tabular}{|c|c|c|c|c|c|c|} 
\cline{2-7} 
\multicolumn{1}{c|}{\multirow{2}{*}{ }} & \multicolumn{2}{c|}{$\ell=2L_o/n_x$} & \multicolumn{2}{c|}{$\ell=3L_o/n_x$} & \multicolumn{2}{c|}{$\ell=4L_o/n_x$}     \\
\cline{2-7} 
\multicolumn{1}{c|}{\multirow{2}{*}{ }} & Toughness &     Percent      &    Toughness    &     Percent      &       Toughness  &       Percent     \\

\multicolumn{1}{c|}{ } &$[J/m^2]$& difference &$[J/m^2]$&  difference   &$[J/m^2]$&  difference   \\
\hline
 Imposed PFF (Corr.) &   2296.6    &      ---            &     2197.2      &      ---           &        2147.8     &      ---             \\
\hline
 CTE $\phi_M=0.80$       &   2309.9    &      0.58\%          &     2247.2      &      2.28\%         &        2222.1     &      3.46\%           \\
\hline
 CTE $\phi_M=0.95$       &   2302.1    &      0.23\%          &     2226.6      &      1.33\%         &        2200.1     &      2.44\%           \\
\hline
\end{tabular}}
\caption{Effective toughness estimates for homogeneous materials with different characteristic length.}
\label{table:3}
\end{table}

It can be seen that CTE with $\phi_M=0.95$ is the best option again for the three lengths but differences in the results between $\phi_M=0.8$ and 0.95 are below $1\%$. Since using CTE values with higher $\phi_M$ is  computationally more demanding in the first time steps, the rest of the calculations of this work will be made with the CTE and $\phi_M=0.8$. 

This accurate results are obtained with the proposed procedure for effective toughness, based on the integration of the loading curve, Eq.\eqref{eq:JaviGc}. This integral, which cannot be obtained using an staggered or similar approach, provides the value of the total energy dissipated, independently on the constitutive and fracture models used. If the PFF model used is not purely variational, the value of the effective toughness computed using the integral of the loading curve or the dissipation functional might differ. The case represented in Fig 4 with $\ell=2L_o/n_x$ and initial CTE crack with $\phi_M=0.80$ is used to quantify this variation. The difference of the value obtained using the dissipation functional with respect to the value obtained in table \ref{table:3} was $4.6\%$.
In the case of a fully variational PFF model, for example the original PFF model without history \cite{miehe2010phase} or other recent approaches \cite{bharali2024micromorphic}, the dissipated energy should coincide with the value of the functional in Eq. \eqref{eq:CrackLenght_ani} and this last value could be used to compute the effective toughness. As an example, the same case in Figure 4 was simulated using a PFF model without history, and the effective toughness obtained using our proposal of integration of the curve Eq.\eqref{eq:SystemDisanyb} and the value obtained using the dissipation potential became identical.
Therefore, since the proposed measure is valid for general cases, the PFF model with history to force irreversibility will be used in the rest of the paper.

\subsubsection{Validation in orthogonal laminates}\label{ssec:lamsVal}
\begin{figure}[htbp]
\centering
\includegraphics[width=150mm]{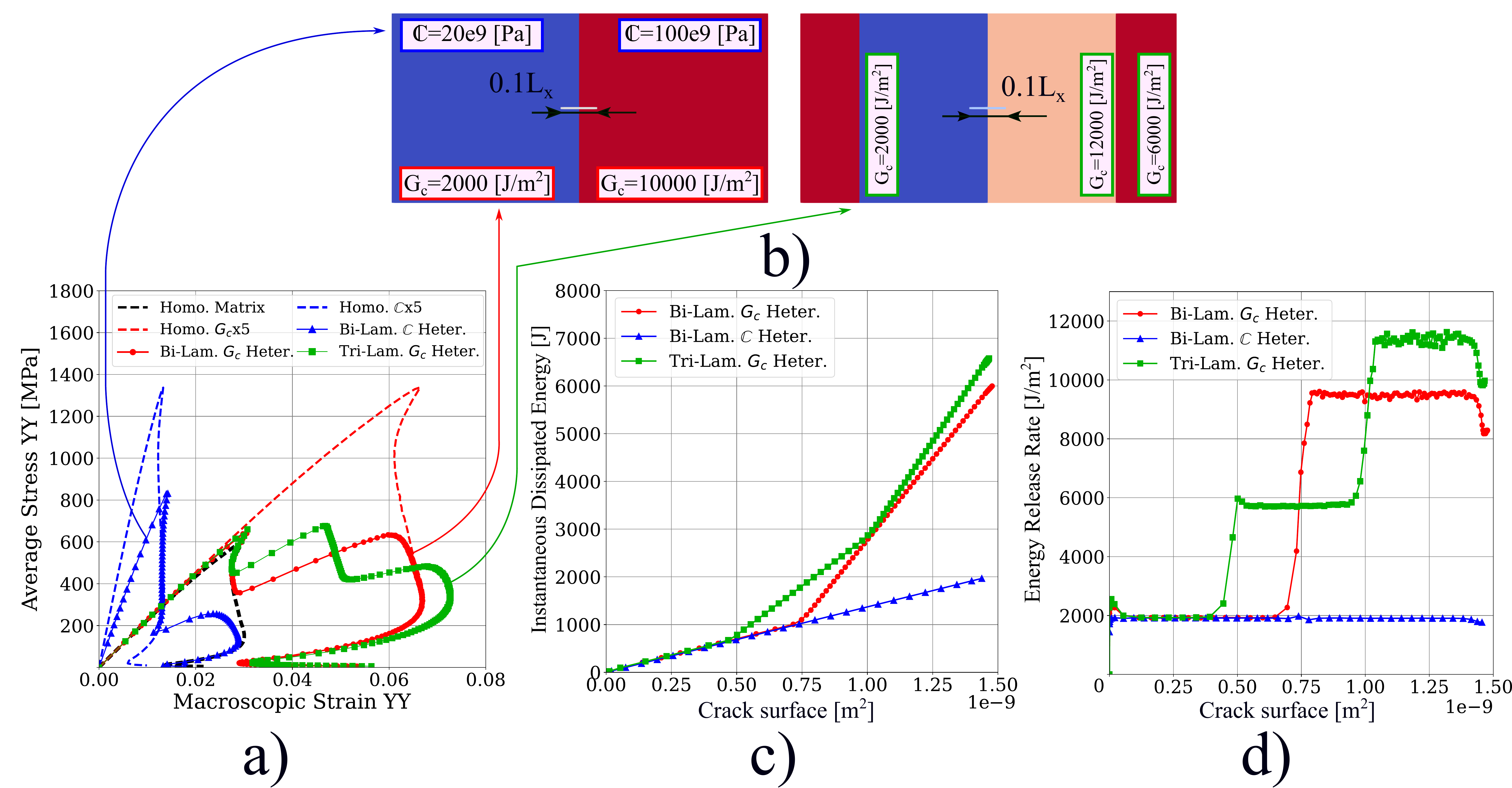}
\caption{\centering{Orthogonal laminates test, for verification of the proposed effective toughness estimate: a) Stress/Strain curves for the three cases in study. b) Schematic of the microstructures analyzed. c) Instantaneous measure of the effective toughness as in Fig.\ref{fig:EffTougnesScheme} for every point. d) Energy release rate measured as in Eq.\eqref{eq:Gmed}.}}
\label{fig:OrtLam}
\end{figure}

To further validate the present proposal for heterogeneous materials, microstructural configurations of orthogonal laminae with initial crack are studied. This type of microstructure was also studied in \cite{schneider2020fft, hossain2014effective} to consider the effect of both local toughness and stiffness heterogeneity. In the former research, it was established that for the geometry depicted in Fig.\ref{fig:OrtLam}b, the effective toughness for a crack perpendicular to the laminates is equivalent to the volumetric average of the local toughness, while, by construction, stiffness heterogeneity played no role. 
In \cite{hossain2014effective} effective toughness is defined with the maximum instantaneous energy release rate, and it is reported that in the case of stiffness heterogeneity the effective toughness increase is similar to the stiffness contrast between sheets, and in the case of toughness heterogeneity the effective metric is equal to the maximum local toughness in the domain. 

The tests have been performed with characteristic length defined by $f_f=2$ and correspond to a rectangular RVE of 185x93x1 voxels and an initial crack modeled with CTE and $0.1L_x$ of initial length. Three cases are studied, two bi-laminates where the first has heterogeneity in toughness and the second local stiffness heterogeneity, and the third case is a tri-laminate with local toughness heterogeneity. Cases with toughness heterogeneity are defined with constant stiffness and cases with stiffness heterogeneity have a constant local toughness, being these constant values defined with the material of the matrix in table \ref{table:1}. In Fig.\ref{fig:OrtLam}c, results of the instantaneous dissipated energy defined in Eq.\eqref{eq:ABC} are depicted similarly as in Fig.\ref{fig:OrtLam}d where values of energy release rate defined in Eq.\eqref{eq:Gmed}.

It can be seen in Fig.\ref{fig:OrtLam}a that several snap-backs are observed in all cases, which correspond to the crack passage between two phases. The homogeneous behavior of the materials used in each case are also depicted for comparison. In Fig.\ref{fig:OrtLam}c the proposed measure applied using the numerical integration in Eq.\eqref{eq:ABC} is shown with respect to the crack surface developed in every time step. In the case with heterogeneous stiffness but same toughness, the evolution of dissipated energy is completely linear, an expected result since the proposed measure is similar to \cite{schneider2020fft}, and for this configuration where a straight crack is developed dissipation is independent on the stiffness of the laminae. This behavior which is the same expected in a homogeneous material can be also observed in Fig.\ref{fig:OrtLam}d, where a constant evolution of energy release rate is obtained. 

In the case where components are heterogeneous in toughness, a piecewise linear evolution is observed, with a slope change every time that surpasses a new phase. This change of slope is due to a higher rate of energy consumption when crack advances in a tougher material. This changes in the rate of dissipated energy are also reflected in Fig.\ref{fig:OrtLam}d, where it can be seen that the energy release rate is constant inside every phase, converging to the local of toughness with an error below $5\%$, presenting jumps when crossing an interface.

In table \ref{table:4} a comparison of the results obtained with our approach with the values obtained with the criteria proposed in \cite{schneider2020fft} and \cite{hossain2014effective} (both summarized in section \ref{ssec:effective_Gc}) in equivalent cases are made. Note that, as discussed in previous section, the results of toughness obtained with the present method in table \ref{table:4} include the correction of the discretization following \cite{bourdin2008variational,aranda2025crack}.

\begin{table}[h!]
\centering
\begin{tabular}{|c|c|c|c|} 
\cline{2-4} 
\multicolumn{1}{c|}{\multirow{2}{*}{ }} &Eff. toughness &   Eff. toughness    &     Eff. toughness    \\
\multicolumn{1}{c|}{\multirow{2}{*}{ }} & as in \cite{schneider2020fft}& as in \cite{hossain2014effective}  &    Proposed  \\
\hline
Bi-Laminate ($G_c$ het.)                &     6000.0    &    10000    &    6040.1   \\ 
\hline
Bi-Laminate ($\mathbb{C}$ het.)         &     2000.0    & $\sim$10000 &    2007.4   \\
\hline
Tri-Laminate ($G_c$ het.)               &     6680.6    &    12000    &    6630.9   \\
\hline
\end{tabular}
\caption{effective toughness estimates for homogeneous materials. All toughness values in $[J/m^2]$.}
\label{table:4}
\end{table}

A very close agreement with the  measure proposed in   \cite{schneider2020fft} is observed in all cases, with differences below $1\%$. On the contrary, the comparison with the results obtained with the metric established as in \cite{hossain2014effective} presents large differences. Regarding heterogeneous stiffness, our proposed measure gives results very similar to the homogeneous case, which means effective toughness is not affected in this case by the heterogeneity in stiffness, even being formulated with the area under the curve of the stress/strain behavior. Since the crack path obtained is straight, no additional toughening mechanism exists in the case of heterogeneous stiffness, so the measure resembles the one in \cite{schneider2020fft}. On the contrary, using the metric by \cite{hossain2014effective}, based on the maximum of the instantaneous release rate, results in an increase in the effective toughness even for a straight crack. Regarding heterogeneous fracture energy cases, both approaches in the literature predict an effect of local toughness on the effective value. Again, since the crack path predicted is straight, our estimate results identical to the one of  \cite{schneider2020fft}, while the value using the definition in \cite{hossain2014effective} would provide an effective toughness equal to the lamina with maximum fracture energy.

\subsection{Effective toughness estimates in composites}\label{ssec:EffToughval}

\subsubsection{Single fiber: stiffness vs toughness heterogeneity effect}\label{ssec:1inc}
\begin{figure}[htbp]
\centering
\includegraphics[width=120mm]{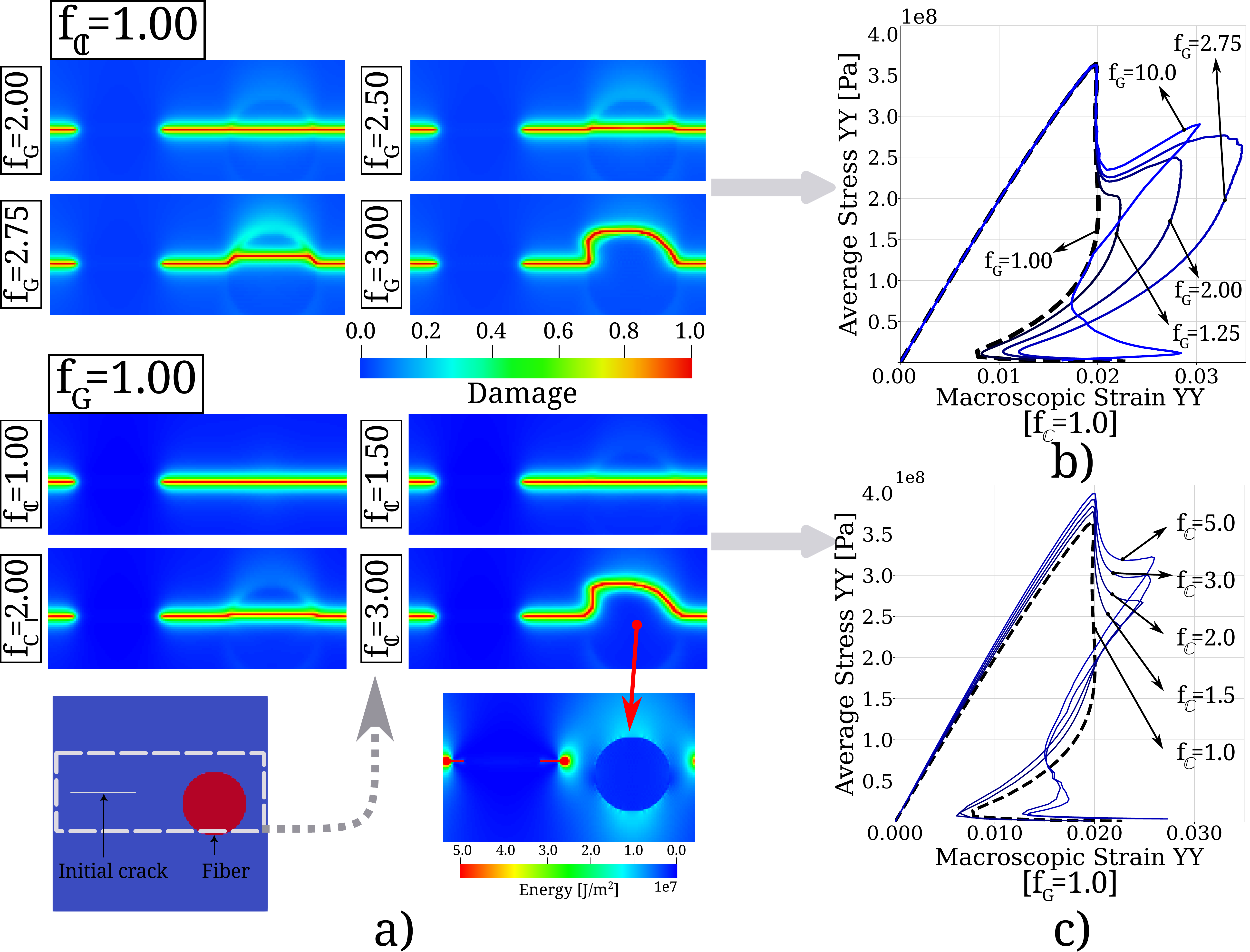}
\caption{\centering{Single fiber simulations: a) Damage fields for different $f_{G}$ and $f_{\mathbb{C}}$. b) Stress/strain curves evolution for $f_{\mathbb{C}}=1$. c) Stress/strain curves evolution for $f_{G}=1$.}}
\label{fig:singleFiber}
\end{figure}

A 2D periodic RVE with a single fiber embedded in a homogeneous matrix is studied here aiming at understanding the effect of interfaces non-orthogonal to the crack path. The proposed RVE is shown in Fig.\ref{fig:singleFiber}a. Different values of fiber stiffness and local toughness are selected, having a ratio with respect to the corresponding property in the matrix of $f_{\mathbb{C}}$ and $f_{G}$ respectively.

The simulations consider an initial crack of $0.3L_x$ with the CTE defined in section \ref{ssec:IniCrack} and spanning the factors defining matrix properties over the ranges $0.5\le f_{\mathbb{C}}\le 5$ and $1\le f_{G}\le 10$. The obtained stress/strain curves are depicted in Figs.\ref{fig:singleFiber}b and Fig.\ref{fig:singleFiber}c, together with selected plots of damage distributions in Fig.\ref{fig:singleFiber}a.

Two extreme cases are analyzed first, materials having only heterogeneity in toughness and materials presenting only stiffness heterogeneity. In Fig.\ref{fig:singleFiber}b the stress-strain curves obtained for the first case are presented, microscopic homogeneous elastic properties $f_{\mathbb{C}}=1$ and different values of the ratio between toughness of the phases. All curves are coincident with the homogeneous case (in black dashed line) until the crack reaches the fiber, where the behavior spreads depending on the level of heterogeneity. If $f_G$ is low, the crack has the tendency to cross the interphase and the effective toughness increases with the local toughness contrast between the fiber and the matrix. This is reflected in the stress-strain curve as an increase of its area under the curve, Fig.\ref{fig:singleFiber}b. This increase in toughness is gradual, starting to a lesser extent for $f_G=2.50$, where a diffuse damage around the fiber starts to appear that grows until it drives the crack to avoid the fiber. For ratios larger than  $f_G=3.0$, the crack fully deflects and bypasses the fiber as depicted in Fig.\ref{fig:singleFiber}a because it becomes energetically more convenient to cross a longer path on a less tough material. This crack behavior agrees with \cite{schneider2020fft}, since a gradual change from crossing to bypassing the fiber is observed for increase in toughness ratio. Also \cite{hossain2014effective} definition of effective toughness accounts for this effect.

The other case of interest is materials with homogeneous fracture energy, $f_G=1.0$, and different local stiffness. In Fig.\ref{fig:singleFiber}a it can be observed that for small values of $f_\mathbb{C}$ the crack penetrates the fiber following a straight path, cases in which no increase in energy (area behind the curve) or effective toughness is found. For higher values of $f_\mathbb{C}$, the strain concentrates around the fiber due to the difference in stiffness, and the energy inside the fiber becomes insufficient to generate any damage, as seen at the bottom of Fig.\ref{fig:singleFiber}a, which leads the crack to bypass the fiber. 
This case illustrates how, despite the microscopic strain fields do not influence the energy dissipated, their heterogeneous distribution can lead to a crack deflection making stiffness heterogeneity a toughening mechanism. This effect is also reported in \cite{hossain2014effective} where the effect of crack deflection due to stiffness local contrast is reported. On the contrary, the definition of effective toughness in \cite{schneider2020fft} does not account for this toughening mechanism since a uniform effective elastic response is assumed at the microscale. The stiffness heterogeneity has an effect on the stress-strain curves, as shown in Fig. \ref{fig:singleFiber}c, which deviate from the homogeneous case. In cases where the crack penetrates the fibers, the curve alternates between increasing and decreasing areas, compensating the total energy consumed which results almost equivalent to the homogeneous case. On the contrary,  when the fiber deflects the crack, there is a net increase in total area under the curve indicating more consumed energy and therefore higher effective toughness. 
 
\begin{figure}[htbp!]
\centering
\includegraphics[width=150mm]{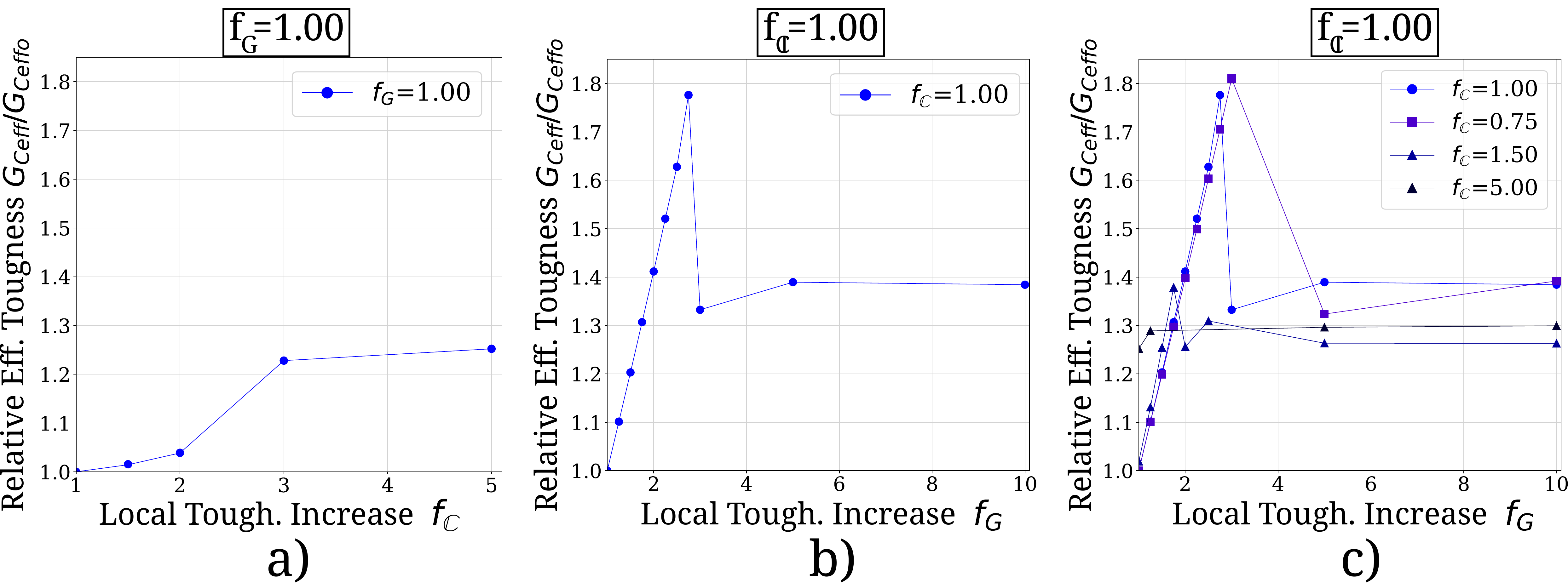}
\caption{\centering{Single fiber simulations: a)effective toughness estimate for all cases with $f_{G}=1$. b)effective toughness estimate for all cases with $f_{\mathbb{C}}=1$. c) Effective toughness estimates for all cases depicted.}}
\label{fig:singleFiber2}
\end{figure}

The evolution of effective toughness as a function of the heterogeneity level between fiber and matrix is represented in Fig.\ref{fig:singleFiber2} for different combinations of stiffness and toughness heterogeneities.

Finally, the combined effect of having both stiffness and toughness heterogeneity is represented in  Fig.\ref{fig:singleFiber2}c (including the case in Fig.\ref{fig:singleFiber2}b for comparison). If the fiber is stiffer than the matrix, $f_\mathbb{C}>1$, the bypassing mechanism appears for a smaller ratio in local toughness. The toughening for sufficiently large ratios of $f_G$ and $f_\mathbb{C}>1$ is in the order of the one reached for a homogeneous elastic composite (between 30-40\% increase with respect to the matrix) without a clear tendency.  Finally, if the fiber is more compliant than the matrix, the opposite effect is observed, making it easier for the crack to penetrate into the fiber and slightly increasing the maximum toughening reachable.

\begin{figure}[h]
\centering
\includegraphics[width=\textwidth]{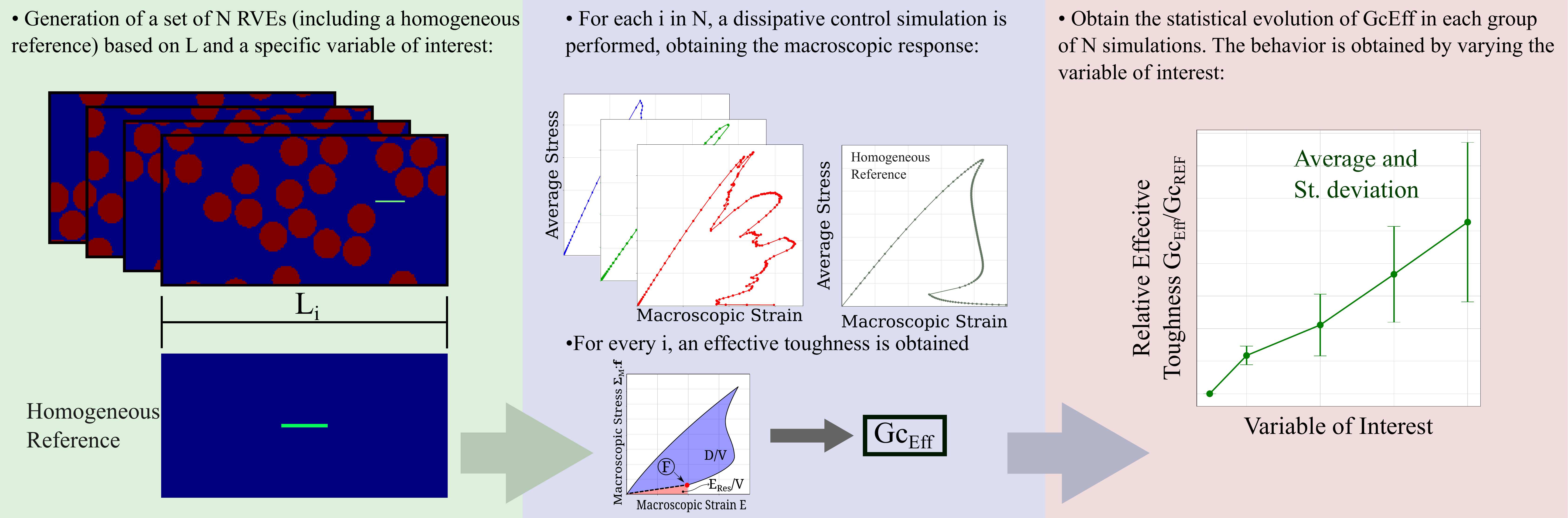}
\caption{\centering{Flow chart of the procedure used to obtain the evolution of the effective toughness in groups of statistically similar geometry.}}
\label{fig:FlowChart}
\end{figure}

\subsubsection{Random fiber reinforced composites}\label{ssec:fibers}

The cross section of composites with random arrays of monodisperse cylindrical fibers is analyzed in this section. The objective is to identify the trend in the evolution of effective toughness across statistically similar groups of fiber arrays, using the method described in section \ref{sec:computTough} to compute the effective $G_c$ in each case. The procedure is illustrated in Fig.\ref{fig:FlowChart}.

\begin{figure}[htbp]
\centering
\includegraphics[width=130mm]{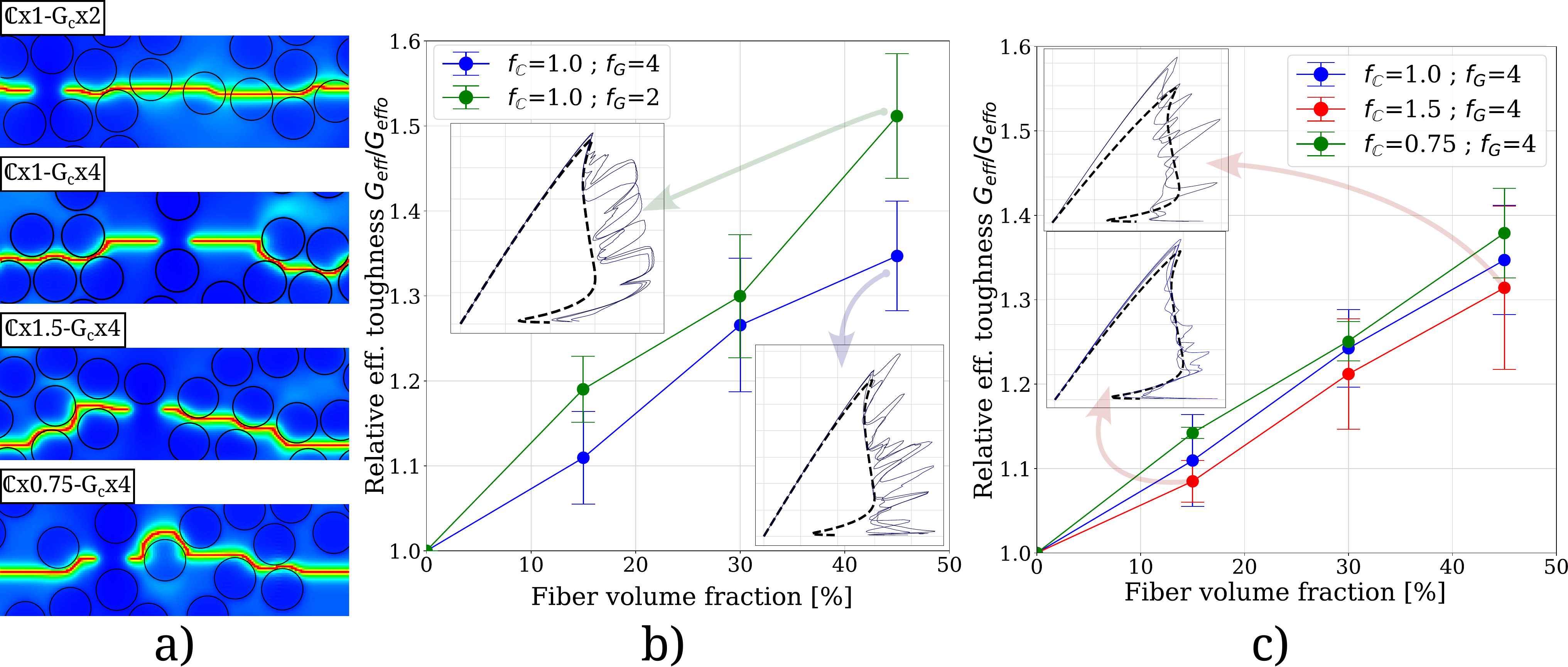}
\caption{\centering{Fiber composite simulations: a) Crack path for different study groups. b) Effective toughness estimates in cases with constant (non-enhanced) local stiffness. c) Effective toughness estimates in cases with constant local toughness (enhanced x4)}}
\label{fig:Fibers}
\end{figure}

The RVEs used are quadrilateral and contain 20 fibers and volume fractions $Fvf=[0\%, 15\%, 30\%, 45\%]$. A CTE initial crack of $0.1L_x$ of length is used in all the simulations. As in the previous section, both stiffness and toughness heterogeneity between the fibers and the matrix will be considered, characterized again with the ratios $f_G$ and $f_\mathbb{C}$ respectively. The results obtained are represented in Fig.\ref{fig:Fibers}.

First, composites with homogeneous stiffness but heterogeneous toughness are considered, with $f_G=4$ and $f_G=2$. These values are selected motivated by the results obtained for the single fiber, in which the phenomenon of fiber penetration or crack deflection was observed respectively for ratios below and above $f_G=2.75$ for $f_\mathbb{C}=1$. To capture statistical effects, for each volume fraction 7 individual cases have been simulated. In  Fig.\ref{fig:Fibers}b the average effective toughness obtained including the standard deviation bars is represented as a function of the volume fraction. The results are normalized with the effective toughness of the homogeneous material. In the two cases, a monotonic increase of the effective toughness is obtained, similar to the study in \cite{aranda2025crack}, obtaining for $Fvf=45\%$ increases close to $30\%$ for $f_G=4$ and close to $50\%$ for $f_G=2$. Similar standard deviations for every case are observed, below $8\%$. The toughness reached for $f_G=2$ is higher than for $f_G=4$ in all the cases, as was also observed for the single fiber. The origin of this tougher behavior is the change of toughening mechanism from fiber crossing, where toughness is increased by the higher energy released within the fiber, to fiber bypass, where toughening occurs by the increase of crack length through the matrix. In Fig.\ref{fig:Fibers}a both mechanisms are observed for $f_G=2$ while for $f_G=4$ no penetration is observed. These mechanisms are also reflected in the resulting stress-strain curves, represented in Fig.\ref{fig:Fibers}b for two cases with $Fvf=45\%$. Crack deflection is perceived in those curves with localized sharp peaks for every change of direction. 
On the contrary, when crack penetration occurs the peaks are less sharp and produce a higher increase of the consumed energy, effect that can only be observed for $f_G=2$.

The second group of simulations corresponds to a material with a fixed toughness heterogeneity and three levels of stiffness heterogeneity in the range $f_\mathbb{C}=[0.75, 1.5]$. An effective toughness increase with volume fraction is also observed (Fig.\ref{fig:Fibers}c), where the three materials present a very similar tendency due to the equivalence of toughening mechanisms, fiber bypass in all the cases. The microscopic crack configurations reached, represented in the last two damage maps of Fig.\ref{fig:Fibers}a, clearly show this unique mechanism. Among the different stiffness ratios, a small increase of the effective toughening is observed with the fiber compliance. As in the case with a single fiber, this can be explained by the strain distribution around the fibers, causing less fiber penetration in the case of high stiffness and more penetration in cases with less stiffness. Regarding stress-strain curves, peak shape observed corresponds to fiber bypass, where the peaks formed in the case with $15\%$ volume fraction of fiber are smaller than for $40\%$, due to the smaller radius of the fibers.

\subsection{Toughening on elastic polycrystals}\label{ssec:resElasPolyX}

In this section, the effective toughness of 2D elastic polycrystals will be analyzed. Parallel to the case of composites, the material used in this section is formulated considering two types of anisotropy, in stiffness and in the energy release rate. The anisotropy in stiffness considers cubic materials, with three independent elastic properties, and is controlled by the Zener ratio, $z$, as
\begin{equation}\label{eq:zener}
z=\frac{2C_{1212}}{C_{1111}-C_{1122}}
\end{equation}
where $z=1$ corresponds to an isotropic Hookean material. The anisotropy in the energy release is set to resemble cleavage, considering a single weak crystal direction tensor, characterized by the tensor $\boldsymbol{A}$ of Eq.\eqref{eq:PFFani}. The level of anisotropy in this case is controlled with the parameter $\beta$, defined in Eq.\eqref{eq:PFFani}. The toughening mechanism in a polycrystal is crack deflection, which will depend on the anisotropy parameters already mentioned. Regarding the effect of microstructure, texture or local misorientation will also control the deflection and therefore the toughening, but these effects are not considered in this study and random orientations are assumed. It is noted that the absolute grain size should not affect the results since our approach is based on homogenization and scale separation \cite{schneider2020fft,ernesti2022computing}.

The domain used is a rectangular RVE of 257x129x1 voxels with a similar characteristic length than previous sections in which grains are modeled using a weighted Voronoi tessellation, resulting in a log-normal grain size distribution \cite{lucarini2021fft}, using for this study a standard deviation of $0.25\%$ of the average grain size and an average of 60 grains in every RVE (unless otherwise stated). Grain orientations are chosen randomly in three dimensions (3 Euler angles) and stiffness is rotated accordingly. In the case of the anisotropy tensor for cleavage, $\mathbf{A}$, it is obtained by rotation and projection of the normal of the weak fracture plane on the RVE cross section. Initial cracks are introduced with the same CTE definition of previous sections and the stiffness tensor constants required for defining the unrotated state $\mathbb{C}_{o}$ in Eq.\eqref{eq:Crot} are corresponding with the matrix material of table \ref{table:1}, being $C_{1111}=24$ GPa, $C_{1122}=8$ GPa, while $C_{1212}$ is obtained as a function of the Zener ratio $z$ following Eq. \eqref{eq:zener}.

\subsubsection*{Effect of elastic anisotropy}
\begin{figure}[htbp]
\centering
\includegraphics[width=140mm]{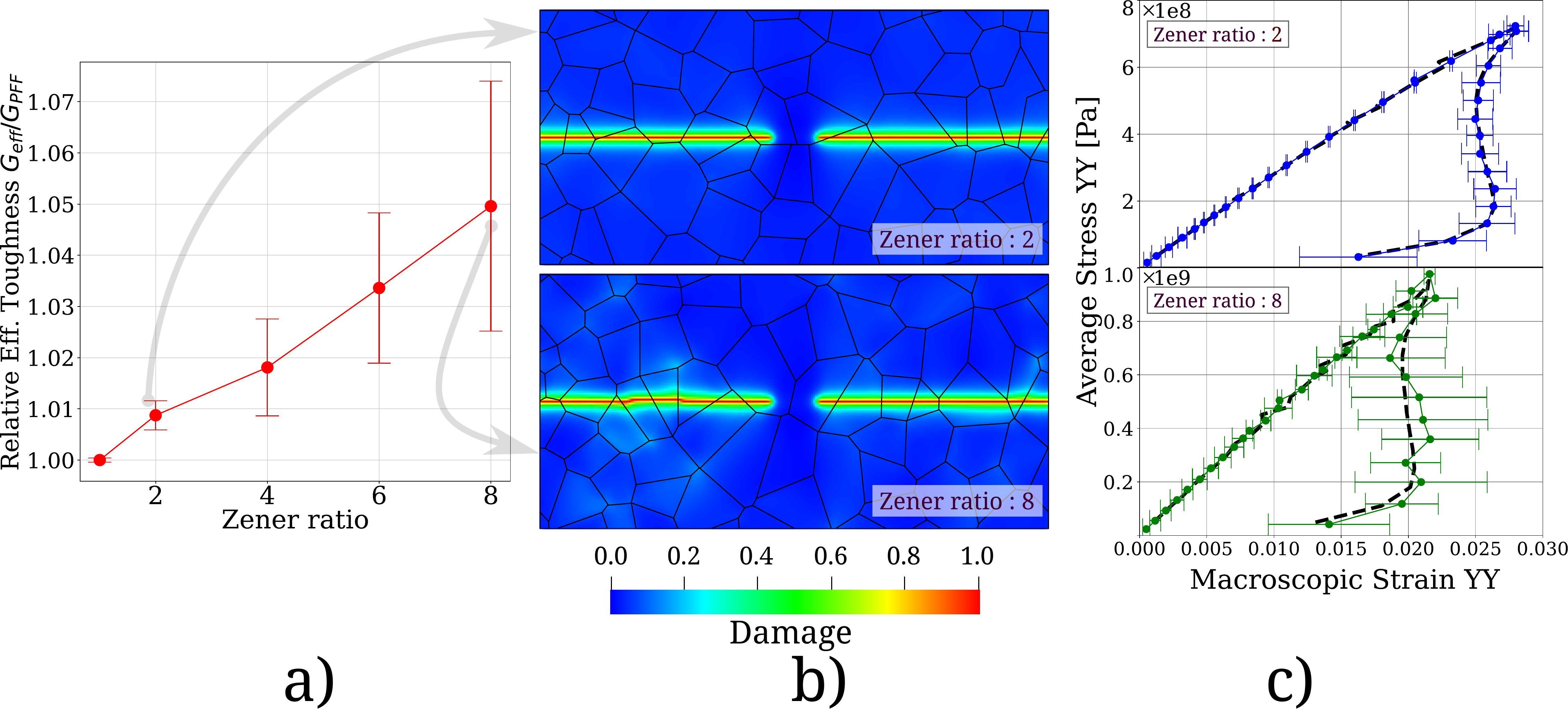}
\caption{\centering{Zener ratio parametric study over polycrystal RVEs: a) Effective toughness estimates for various values of $z$. b) Examples of crack path for the groups with $z=2$ (above) and $z=8$ (below), both with average grain size $30 \mu m$ and $\beta=0$. c) Stress/strain average curves for all groups )}}
\label{fig:PolyX1}
\end{figure}

The first case analyzed is the toughening due to elastic anisotropy, which is controlled by the Zener ratio (Eq. \ref{eq:zener}). The reference case is a homogeneous and isotropic material ($z=1$ and $\beta=0$) in which polycrystal geometry has no effect, and the other cases correspond to Zener values $z=2,4,6,8$, each case including a set of different simulations that vary the grain distribution and keep the number of grains as a constant for all cases. The results of the effective toughness as a function of $z$ are represented in Fig.\ref{fig:PolyX1}a. An example of the resulting crack paths obtained for each Zener ratio is represented in Fig. \ref{fig:PolyX1}b. Finally, Fig.\ref{fig:PolyX1}c shows the average stress/strain curves obtained and used to compute the effective toughness.

The results show an increase in the effective toughness with the level of anisotropy. This increase reaches 5\% for the highest Zener ratio, while toughening for smaller values of this parameter, typical of standard alloys (e.g. very anisotropic crystals as Ni have $z=2.5$ \cite{rosler2007mechanical}), will be below 2\%. The toughening mechanism associated with this increase, crack deflection, is clearly observed in Fig.\ref{fig:PolyX1}b. These deflections are caused by heterogeneous accumulations of elastic energy in the domain caused by the strain incompatibility between grains. These results are similar to those obtained for local stiffness variations in fiber composites, where strain accumulation was also the origin of crack deflection and toughness increase. It's clear from Fig.\ref{fig:PolyX1}b that low heterogeneity as in the case of $z=2$ causes low crack deflection and almost no toughening, which emphasizes the idea that differences in local rigidity do not increase toughness unless it generates a significant crack deflection. Also, dispersed damage is observed all over the domain, due to the strain energy concentration around grain boundaries. 

In Fig.\ref{fig:PolyX1}c, the stress/strain curves of each polycrystal are represented together with the results of an equivalent homogeneous material. The effective stiffness of the equivalent material is obtained, for each Zener ratio, using Fourier-Galerkin homogenization and RVEs with a very large number of grains. The effective toughness is the same as that of the crystals since this case assumes a homogeneous fracture energy.

\subsubsection*{Effect of anisotropy in toughness}
\begin{figure}[htbp]
\centering
\includegraphics[width=140mm]{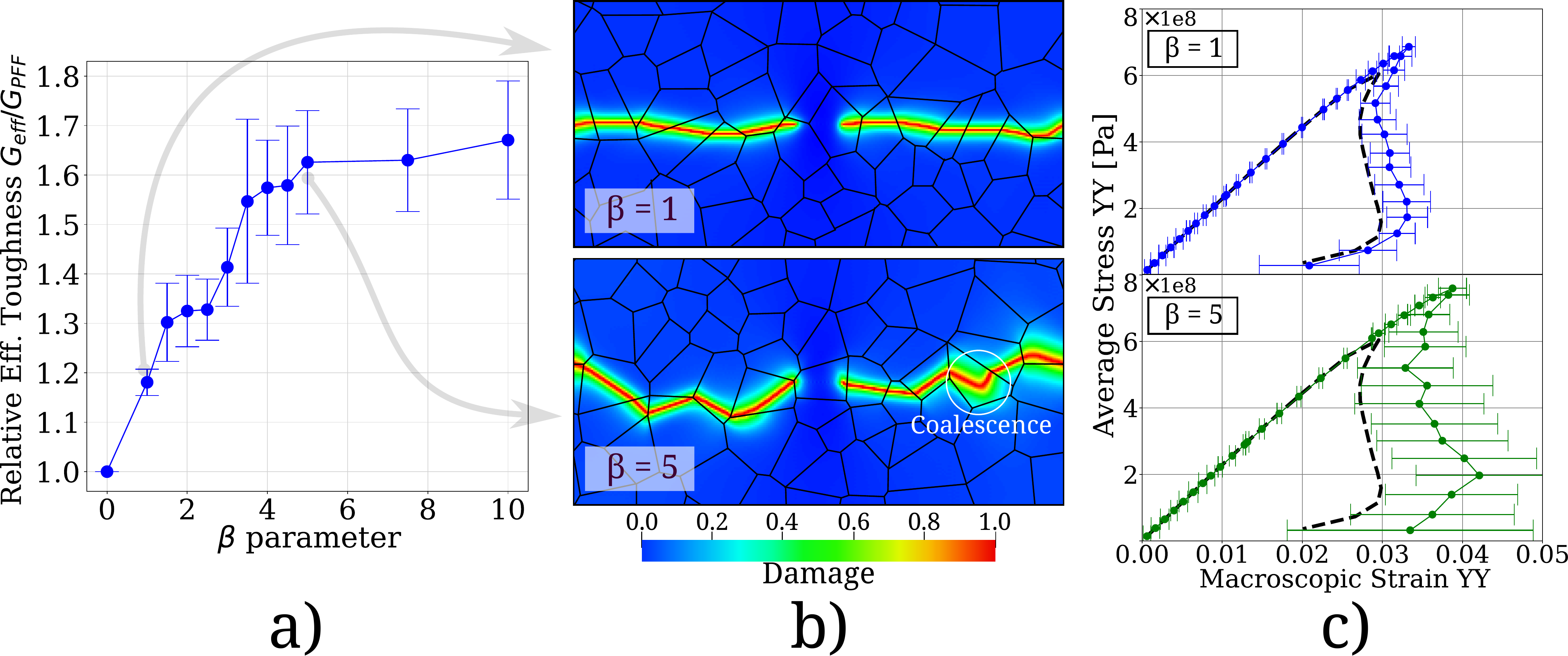}
\caption{\centering{Damage anisotropy parametric study over polycrystal RVEs: a) effective toughness estimates (corrected) for various values of $\beta$. b) Examples of crack path for the groups with $\beta=1$ (above) and $\beta=5$ (below). c) Average stress/strain curves for the entire groups $\beta=1$ (above) and $\beta=5$ (below) compared to the behavior in homogeneous case}}
\label{fig:PolyX2}
\end{figure}

The second case considered is the analysis of the effect of the fracture energy anisotropy. The same polycrystalline RVEs as in the previous section are tested, setting here the Zener ratio at $z=1$ and varying $\beta$ in the range $[1,10]$. The reference toughness is the homogeneous case, which corresponds to $\beta=0$. The resulting effective toughness as a function of $\beta$ is shown in Fig.\ref{fig:PolyX2}a, crack paths in Fig.\ref{fig:PolyX2}b and the stress/strain curves in Fig.\ref{fig:PolyX2}c.

The effective toughness obtained as a function of $\beta$ (Fig. \ref{fig:PolyX2}a) is highly affected by the level of anisotropy in the crystal toughness. The effective value shows a steep increase until reaching $\beta\approx4$, where the effective response saturates. The toughening obtained is high compared with the one observed as a result of the elastic anisotropy, reaching here up to $60\%$. It is interesting to note that the toughness obtained using the proposed approach is very similar to the one obtained in \cite{ernesti2022computing} using the definition of effective toughness in \cite{schneider2020fft}, reaching a saturation of the effective toughness almost identical to the one obtained here.

The microscopic toughening mechanism is the increment of crack length resulting from the constant change of direction of the crack paths. This effect can be observed in  Fig.\ref{fig:PolyX2}b where crack paths show a clear cleavage behavior and the crack propagates in a completely transgranular way, changing its direction sharply every time the crack enters a new grain. This effect is higher with the increase of $\beta$, for $\beta=1$ the crack still propagates horizontally in many grains independently of their orientation, while for $\beta=5$ the crack deviates in each grain to the cleavage plane. The stress/strain behavior of all the polycrystals is identical before crack propagation. After propagation, the stress-strain curves of every case follow a path that is always above the homogeneous behavior. The error bars in the case of $\beta=5$ are also higher than in the case of $\beta=1$, implying the need for larger RVEs to obtain statistically representative results.

\subsubsection*{Effect of the RVE size}
In order to analyze the effect of RVE size (number of grains) statistical RVEs containing ensembles of 5, 8, 18, 30, 60, 140, 460, 1360, 1690, and 2160 grains are analyzed. In this study, the elastic response is chosen as isotropic, $z=1$,  and damage anisotropy corresponds to a value of $\beta=1$. A constant value of $\ell$ is chosen, which might introduce a small effect of the absolute grain size. This effect, which is purely numerical and should not exist in LEFM when $\ell\rightarrow0$, is minimized here by choosing an  $\ell$ sufficiently small with respect to grain size in all the cases considered. 

\begin{figure}[htbp]
\centering
\includegraphics[width=140mm]{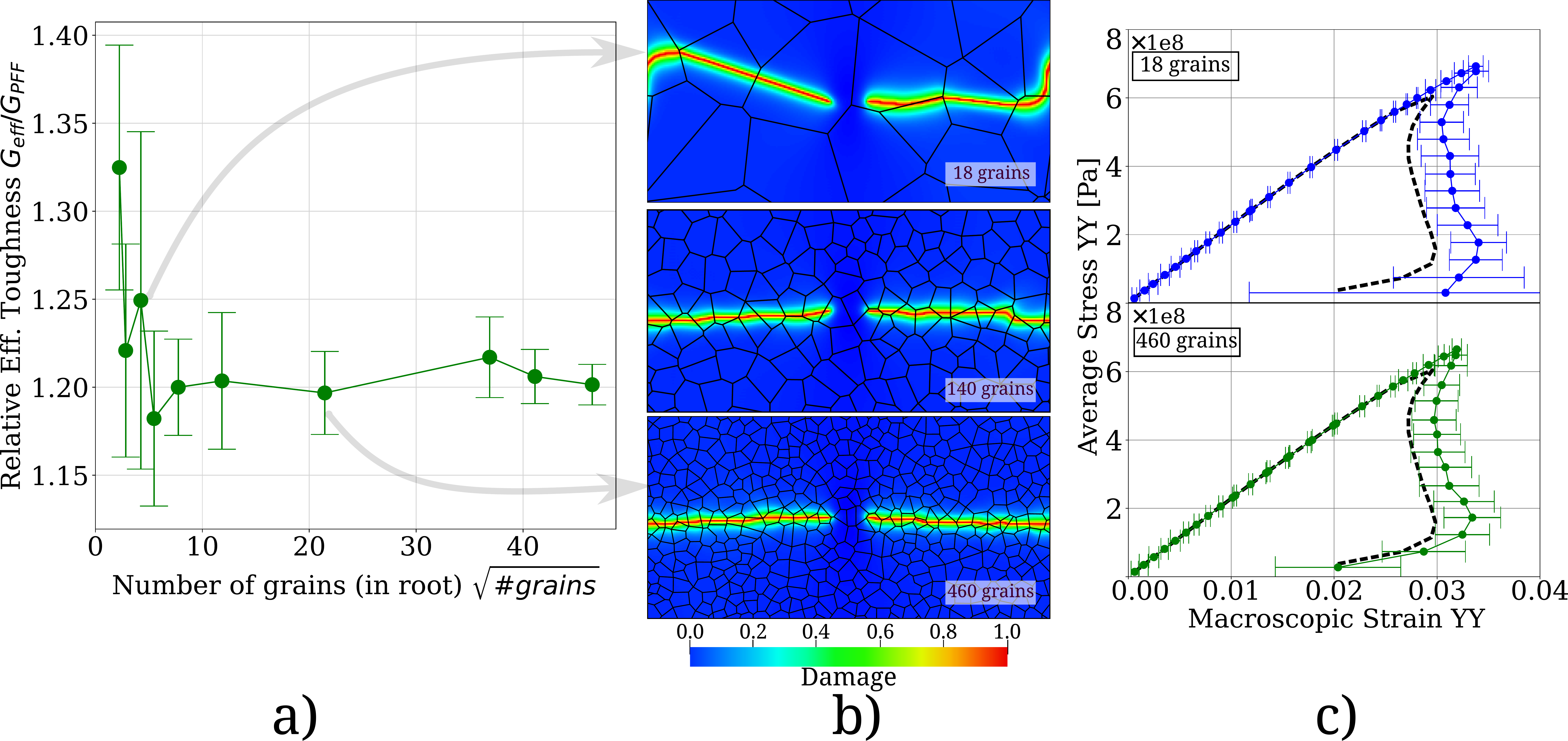}
\caption{\centering{Parametric study on grain average number over polycrystal RVE volume: a) effective toughness estimates for various grain average number. b) Examples of crack path for the groups with grain average number of 18 (above), 140 (middle) and 460 (below) grains. c) Stress/strain curves for the entire groups of grain average number of 18 (above) and 460 (below).}}
\label{fig:PolyX3}
\end{figure}

The results of the effective toughness as a function of the average number of grains per unit length (proportional to the square root of the total number of grains) are represented in Fig.\ref{fig:PolyX3}a. On the same figure, the crack paths for 3 RVE sizes are represented in Fig.\ref{fig:PolyX3}b, and the average stress/strain behavior for the RVEs with 18 and 460 grains are represented in Fig.\ref{fig:PolyX3}c, including statistical dispersion.
In Fig.\ref{fig:PolyX3}a it is observed that the effective toughness shows an increase of $20\%$ to $35\%$ for all the RVE sizes considered and that this value seems to converge for a high number of grains, as shown in  \cite{ernesti2022computing}. As expected, standard deviation also decreases with RVE size, as it happens when homogenizing, for example, the mechanical response.  Regarding the toughening mechanism, as expected, it is independent of the RVE size and always corresponds to crack path changes. Looking at the average stress-strain curves (Fig.\ref{fig:PolyX3}c) it can be observed that the qualitative average response of the polycrystal is very similar to that of a homogeneous material. Only when the number of grains is very small, the coalescence part of the stress-strain curve changes with respect the homogeneous materials. This effect is due to the periodicity and becomes negligible for sufficiently large RVEs.



\section{Conclusions}
An effective toughness estimate is proposed based on PFF-FFT simulations of heterogeneous RVEs performed under stable crack growth control \cite{aranda2025crack}. The definition of the toughness corresponds to the total energy dissipated after the total fracture of the RVE ---which can be accurately obtained thanks to the crack area control---, divided by the RVE area (or length in 2D). The estimate considers the effect of both toughness and elasticity heterogeneity on the overall fracture energy. The proposed framework also considers anisotropic phases, both in elasticity and toughness (cleavage), which result in heterogeneous materials when present in different orientations in the microstructures. PFF in FFT follows previous works of the group \cite{zarzoso2025fft,aranda2025crack} and in order to improve the prediction of toughness, different ways of introducing an initial crack are explored. It is found that crack-tip enrichment, proposed in \cite{singh2016fracture} provides the best results with a reasonable numerical efficiency. The method is applied to obtain the effective toughness of composites and elastic polycrystals in a series of examples.

The estimate proposed here presents some common characteristics with the variational approach developed by Schneider in \cite{schneider2020fft} and the PFF-based approach proposed by Hossein et. al. in \cite{hossain2014effective}. In \cite{schneider2020fft} the effective toughness is obtained by finding the crack path in which the fracture energy dissipated is minimal. In both approaches, the overall toughness converges to the integral of the fracture energy through the resulting crack area in the RVE and almost identical results are obtained for the same microstructural crack paths (as shown in this paper in the case of laminates). However, there are some important differences. In our approach, an initial crack should be introduced in the RVE to avoid artificial crack nucleation, which is rigorously not accounted for in brittle PFF. Although results are almost unaffected by initial crack size, a dependency on the location of the crack is introduced, which influences the crack path that might differ from the minimal energy path. This effect can be overcome by sampling on different crack positions. The second important difference is that the present approach considers the effect of the stiffness heterogeneity, which by construction is not accounted for in \cite{schneider2020fft}. This heterogeneity results in a complex strain distribution at the microscale which can induce crack path changes even in the absence of heterogeneity in local toughness. This toughening effect is captured in the present approach both for composites and polycrystals.

With regards to the metric proposed by Hossein et. al. \cite{hossain2014effective}, our method is also based on PFF simulations and therefore both require the introduction of initial cracks and are able to account for the influence of microscopic elastic heterogeneity. The main difference with Hossein's approach is that here the energy release rate is defined as a function of the total dissipated energy, while in \cite{hossain2014effective}, the maximum instantaneous value is used as an estimate of the toughness. Also, there are differences in the RVE PFF simulations; in \cite{hossain2014effective} crack evolves under \emph{surfing BCs} based on the displacements around a crack, while in our approach, macroscopic far-fields are applied and controlled to produce a constant crack growth rate.   

Focusing on the effective toughness results, in the case of composites, the toughening mechanism found is  crossing phases with different toughness or the deflection of the crack path through the matrix. For elastic heterogeneity but toughness homogeneity, the toughening mechanism consists in the deviation of the crack to avoid fibers, an effect also captured in \cite{hossain2014effective}. This effect saturates and no changes in toughness are obtained above a critical value of stiffness heterogeneity. In the case of elastic homogeneous response and heterogeneity in toughness, fiber crossing is found for relatively small phase contrast, leading to a high increase in toughness. However, after a certain contrast in toughness, the only mechanism observed is crack deflection, and for tougher fibers, the effective response is kept.

In the case of elastic polycrystals, both anisotropy in the elastic response and fracture energy are considered, controlled by the Zener ratio in the first case, and a parameter $\beta$ controlling cleavage fracture. The toughness mechanism in this case is crack deviation, being both types of anisotropy able to achieve it. The anisotropy in the fracture energy shows a great increment in effective toughness (up to $\approx50\%$ for our parameters) and saturation after reaching a critical anisotropy value. Stiffness anisotropy shows a weaker, but non-negligible, toughness increase with a near-linear dependency with the Zener ratio, highlighting that the effect of stiffness variations also has influence in the proposed measure.

The procedure described in this work was applied to simplified 2D cases that in some cases are indeed representative of the full behavior such as the case of fiber-reinforced materials under transverse I-mode loading. The extension to other fracture modes and more complex geometries in three dimensions, including for example real polycrystals, can be done without modifications to the scheme. Future studies will pursue these extensions using a measurement method equivalent to the one presented here.

\section*{Acknowledgements}
PA is grateful for the support of the doctoral program ANID-BECAS CHILE 2020-72210273 from the Government of Chile and to the "Vicerrectoría de Investigación, Desarrollo e Innovación" of the "Universidad de Santiago de Chile". JS acknowledges the support of project TED2021-130255B-C32 from "Ministerio de Ciencia y Educación".

\bibliographystyle{unsrt}
\bibliography{Bibliography}

\end{document}